\documentclass[pre,aps,twocolumn,showpacs,preprintnumbers,superscriptaddress]{revtex4}

\def\draftmode{0}

\usepackage{graphicx}
\usepackage{amssymb}
\usepackage{amsmath}
\usepackage{times}
\usepackage{graphicx}
\usepackage{braket}

\usepackage{todonotes} 
 
\renewcommand{\d}{\mathrm{d}}
\newcommand{\dd}{\Delta}

\newcommand{\vth}{V_\text{th}}
\newcommand{\vreset}{V_\text{r}}

\newcommand{\cv}{C_\text{V}}

\newcommand{\lrrund}[1]{\left( #1 \right)}
\newcommand{\lreckig}[1]{\left[ #1 \right]}

\newcommand{\op}[1]{\operatorname{#1}}

\newcommand{\od}[2]{\frac{\mathrm{d}#1}{\mathrm{d}#2}}

\DeclareMathSymbol{\shortminus}{\mathbin}{AMSa}{"39}

\begin{document}
\title{Low-dimensional firing-rate dynamics for populations of renewal-type spiking neurons}

\author{Bastian Pietras}
\affiliation{Institute of Mathematics, Technical University Berlin, 10623 Berlin, Germany.}
\affiliation{Bernstein Center for Computational Neuroscience Berlin, 10115 Berlin, Germany.}

\author{No\'e Gallice}
\affiliation{Brain Mind Institute, \'Ecole polytechnique f\'ed\'erale de Lausanne (EPFL), Station 15, CH-1015 Lausanne, Switzerland}


\author{Tilo Schwalger}
\affiliation{Institute of Mathematics, Technical University Berlin, 10623 Berlin, Germany.}
\affiliation{Bernstein Center for Computational Neuroscience Berlin, 10115 Berlin, Germany.}

\date{\today}

\begin{abstract}
  The macroscopic dynamics of large populations of neurons can be mathematically analyzed using low-dimensional firing-rate or neural-mass models. However, these models fail to capture spike synchronization effects and non-stationary responses of the population activity to rapidly changing stimuli. Here, we derive low-dimensional firing-rate models for homogeneous populations of neurons modeled as time-dependent renewal processes. The class of renewal neurons includes integrate-and-fire models driven by white noise and has been frequently used to model neuronal refractoriness and spike synchronization dynamics. The derivation is based on an eigenmode expansion of the associated refractory density equation, which generalizes previous spectral methods for Fokker-Planck equations to arbitrary renewal models. We find a simple relation between the eigenvalues characterizing the time scales of the firing rate dynamics and the Laplace transform of the interspike interval density, for which explicit expressions are available for many renewal models. Retaining only the first eigenmode yields already a decent low-dimensional approximation of the firing-rate dynamics that captures spike synchronization effects and fast transient dynamics at stimulus onset. We explicitly demonstrate the validity of our model for a large homogeneous population of Poisson neurons with absolute refractoriness, and other renewal models that admit an explicit analytical calculation of the eigenvalues. The here presented eigenmode expansion provides a systematic framework for novel firing-rate models in computational neuroscience based on spiking neuron dynamics with refractoriness.
\end{abstract}
\maketitle


\section{Introduction}


One of the most successful models in computational neuroscience is the firing-rate model introduced by Wilson and Cowan \cite{WilCow72}. Firing-rate or neural-mass models describe the coarse-grained activity of large populations of neurons and are widely used to model cortical computations \cite{WimNyk14,PerBru18,BenBar95,RubVan15,ShpMor09,WonWan06,SusAbb09} and macroscopic neural imaging data \cite{MorPin13,DipRan18}. Their simplicity in form of first-order differential equations permits a mathematical analysis and a mechanistic understanding of cortical dynamics \cite{WilCow72,JerRox17}.
However, classical firing-rate (FR) models are heuristic models with important limitations.
Although FR models can capture the stationary mean activity, they do not correctly reproduce the \emph{dynamics} of a population of spiking neurons such as transient population activity.
After the onset of a step stimulus, the relaxation of the trial- or population-averaged activity to the stationary state follows a simple exponential time course in FR models, whereas experimental data \cite{TchMal11} as well as spiking neuron models \cite{MatGiu02,GerKis02,SchOst13,GerKis14,Chi17,SchDeg17} exhibit a much richer relaxation dynamics.
Furthermore, even in cases where an exponential approximation is justified, a simple relationship between the relaxation time constant and the underlying neuronal dynamics is not evident \cite{ErmTer10,FouBru02}.
On the other hand, single neuron dynamics crucially influence the properties of large neural networks such as synchronization phenomena \cite{GolRin94,BruHak99,Bru00,ErmTer10,DevRox17,PieDev19} or signal transmission properties via the shaping of network noise \cite{MusGer19}.
Heuristic FR models with first-order dynamics are not suitable to study these effects.

Classical FR models for the population activity can be strictly derived from an underlying microscopic model in the special case where individual neurons are modeled as \emph{non-homogeneous Poisson processes} with first-order rate dynamics \cite{SchChi19}.
Such Poisson models do not exhibit any memory of past spikes, and therefore lack basic properties of neural dynamics such as refractoriness and spike-synchronization.
In contrast, these properties can be readily obtained with neurons modeled as \emph{non-homogeneous renewal processes}, i.e. stochastic spiking neuron models
that keep a memory up to the last spike \cite{Ger00}.
Renewal models are widely used in computational neuroscience:
They encompass mechanistic models such as one-dimensional integrate-and-fire models with white current noise \cite{Bru00,LinDoi05,Ric08,FarVre15} and the spike-response model with stochastic spike-generation \cite{Ger00} as well as more abstract renewal models defined by a hazard rate \cite{GerKis14,SchDeg17}. However, the population rate dynamics for general renewal models is considerably more complicated than classical FR dynamics. For instance, the population rate dynamics can often be formulated as a partial differential equations with non-local boundary conditions (McKean-Vlasov equations). Examples are the Fokker-Planck equation for the density of membrane potentials \cite{AbbVre93,AmiBru97,GerKis14} or the refractory density equation (or, equivalently, the renewal integral equation) for the density of last spike times (age-structured population dynamics) \cite{Ger95,Ger00,ChiGra07,ChiGra08,PakPer09,DumPay17,SchDeg17}. Thus, formally, the population rate dynamics of renewal models is infinite dimensional, reflecting the continuum of internal neuronal states.

In this paper, we put forward a systematic reduction of the population rate dynamics for renewal models to low-dimensional dynamics in the spirit of classical FR models. The reduction is based on the eigenfunction expansion method that has been originally proposed for noisy integrate-and-fire models governed by a one-dimensional Fokker-Planck equation \cite{KniMan96,Kni00,MatGiu02,DoiRin06,GigMat07,SchOst13,AugLad17}. Here, we generalize this approach by using the refractory density equation that holds for arbitrary renewal processes. The refractory density approach is advantageous in several respects: first, it operates on an effectively one-dimensional state space even for multi-dimensional neuron models such as conductance-based models \cite{ChiGra07,ChiGra08}; second, it can be extended to generalized integrate-and-fire models with adaptation \cite{PozMen15,TeeIye18,SchChi19} via a quasi-renewal approximation \cite{NauGer12}; and third, it admits a generalization to finite-size (mesoscopic) populations \cite{SchDeg17}. We will not go into these potential extensions here but will first consider the important and non-trivial case of renewal models and infinitely large (macroscopic) populations. This case allows us to uncover a surprisingly simple relation between the dominant time scales of the population dynamics and basic characteristics of renewal processes. Moreover, the low-dimensional FR dynamics, including the characteristic time scales of the system, can be calculated analytically for simple renewal models, which has been infeasible in the Fokker-Planck framework.

The paper is organized as follows: We first introduce the microscopic model of interacting renewal processes and the corresponding macroscopic population dynamics governed by the refractory density equation (Sec.~\ref{sec:time-depend-renew}). In Sec.~\ref{sec:spectr-decomp}, we briefly review the spectral decomposition method. 
The main part of the theory is presented in Sec.~\ref{sec:eigenv-eigenf-refr}, where the spectral decomposition is applied to the refractory density equation. In particular, we find a characteristic equation for the eigenvalues and calculate the eigenfunctions in terms of the hazard rate, the interspike interval density, or the survival function of the renewal processes. Truncating the infinite modal expansion at a given order allows us to systematically obtain low-dimensional approximations of the FR dynamics in terms of single neuron properties (Sec.~\ref{sec:low-dimens-firing}). 
In Sec.~\ref{sec:spec_decomp_specific}, the theory is applied to different renewal models that permit explicit formulas for the eigenvalue spectrum. We also provide a general expression for the eigenvalue spectrum in terms of the rate and coefficient of variation (CV) of the interspike intervals (ISIs) for the case when the ISI density is approximately Gaussian. The analytical formulas allow us to study the model dependence and structure of eigenvalues and eigenfunctions. The quality of the low-dimensional approximations is assessed in Sec.~\ref{sec:low-dim_pop_dyn} with respect to fast transient dynamics, time-dependent response to stimuli and oscillations (partial synchronization) in recurrently-connected networks. Finally, possible applications and extensions as well as open technical issues are discussed in Sec.~\ref{sec:discussion}. In the Appendix we provide detailed and alternative derivations and show that the eigenvalue spectrum of the refractory density equation and the Fokker-Planck equation is equivalent (Sec.~\ref{sec:equiv-fokk-planck}).

\section{Time-dependent renewal model}
\label{sec:time-depend-renew}

\subsection{Microscopic model}

\begin{figure*}[t]
\begin{center}
\includegraphics[scale=1]{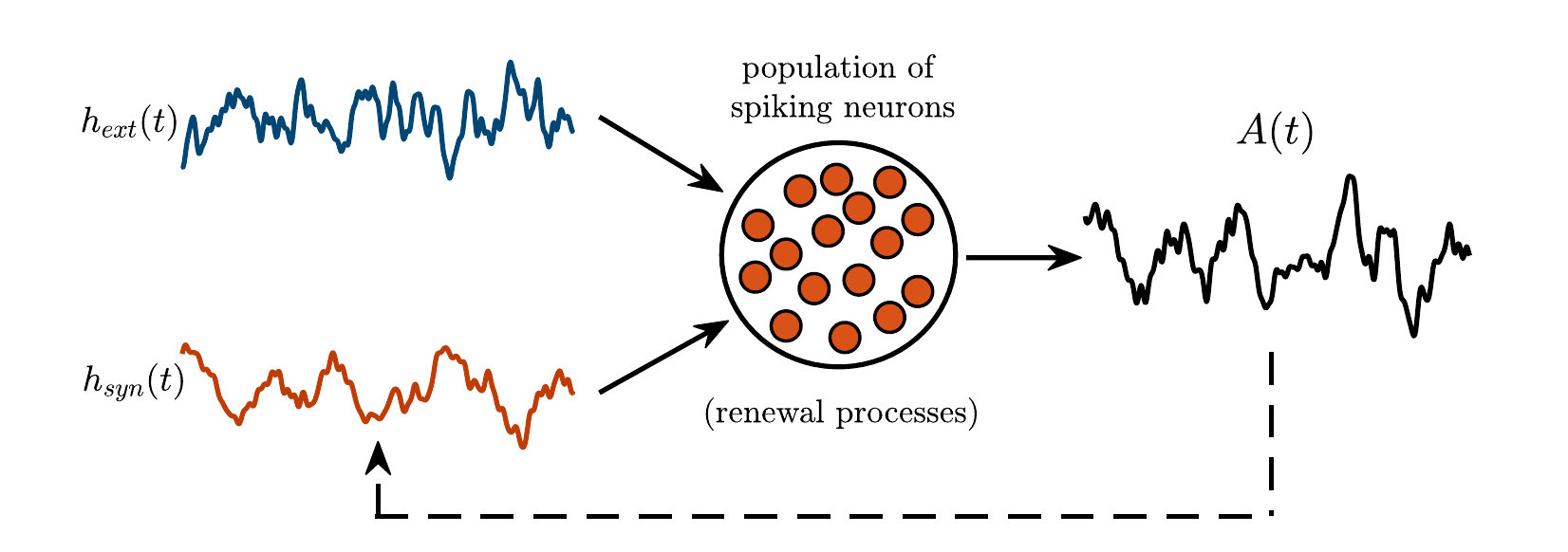}
\caption{Population activity $A(t)$ of spiking neurons that are modeled as time-dependent renewal processes subject to external and recurrent synaptic input currents $h_{ext}(t)$ and $h_{syn}(t)$, respectively.}
\label{fig:1}
\end{center}
\end{figure*}

To keep the arguments simple, we consider a single homogeneous population of $N$ globally coupled neurons. Each neuron is characterized by a variable  $\tau_i(t)$ ($i=1,\dotsc,N$), called ``age'', that describes the time elapsed since its last spike before the present time $t$. 
If, for example, neuron $i$ fired its last spike at time $\hat t_i$, then its age at time $t$ is $\tau_i(t) = t - \hat t_i$.
Spikes of a given neuron $i$ are generated stochastically by a hazard rate $\rho_i(t)=\rho\bigl(\tau_i,h(t)\bigr)$ that only depends on its age $\tau_i(t)$ and a global parameter $h(t)$ common to all neurons. 
The instantaneous probability that neuron $i$ fires in a small time interval $[t,t+\Delta t)$ is hence given by
\begin{equation}
  \label{eq:hazard}
  \text{Pr}(\text{spike in $[t,t+\dd t)$}|\tau_i(t),h(t))=\rho_i(t)\Delta t+o(\dd t)
\end{equation}
for $\Delta t\rightarrow 0$. Mathematically, this class of neuron models corresponds to a time-dependent (or non-homogeneous) renewal process \cite{Ger95,Ger00,GalLoe15_arxiv} or ``age-structured'' process \cite{PakPer09}. 

For simplicity, we will focus on a scalar parameter $h(t)$; a generalization of the theory to multi-dimensional parameters $\vec{h}(t)=\{h_1(t),h_2(t),\dotsc\}$ is straightforward (see Appendix, Sec.~\ref{sec:append-a:-deriv}). 
Multi-dimensional parameters could represent different synaptic input streams in a multi-population setup, but also additional gating variables \citep{ChiGra07} or a slow adaptation \citep{GigMat07} or depression \citep{GigDec15,SchGer18arxiv} variable. In this paper, the scalar parameter  is intended to represent the total input  resulting from external and synaptic inputs (Fig.~\ref{fig:1}):
\begin{align}
  \label{eq:def-filters}
  h(t)&=h_\text{ext}(t)+h_\text{syn}(t),\\
  h_\text{ext}(t)&=\int_0^\infty \kappa(s) I(t-s)\dd s,\\
  h_\text{syn}(t)&=\frac{J}{N}\sum_{j=1}^N\int_0^\infty \epsilon(t')s_j(t-t')\d t'
\end{align}
Here, $\kappa(t)$ and $\epsilon(t)$ are filter functions that describe the effect of a global external input current $I(t)$ and presynaptic spike trains $s_j(t)=\sum_{k}\delta(t-t_k^j)$, where $\delta(\cdot)$ denotes the Dirac delta function and $t_k^j$ is the $k$-th spike time of neuron $j$. The parameter $J$ denotes the synaptic weight. 

Furthermore, we assume that the hazard function $\rho(\tau,h) \geq 0$ is non-negative and vanishes for strong hyperpolarization, $h\rightarrow -\infty$.
Due to the explicit dependence on $\tau$, the hazard function describes the time-course of recovery from firing and can therefore account for neuronal refractoriness. Alternatively, the renewal models can be characterized in terms of the \emph{survival function} 
\begin{equation}
\label{eq:survprob}
S(\tau,h) = \exp\!\left( - \int_0^\tau \rho(s,h) \d s\right),
\end{equation}
or the \emph{interspike interval (ISI) density} 
\begin{equation}
\label{eq:ISI_dens}
P(\tau,h) =\rho(\tau,h) S(\tau,h).
\end{equation}
Given the interdependence of hazard rate, survival probability, and ISI density \cite{Cox62,GerKis14}, one of these functions suffices to fully determine the renewal process. 

%

\subsection{Macroscopic dynamics: refractory density equation}
Our goal is to describe the dynamics of the \emph{population activity} $A(t)$, i.e. the fraction of neurons that fire per unit time, in the limit of an infinitely large population ($N\rightarrow\infty$). More precisely, if $n(t,t+\Delta t)$ denotes the total number of spikes generated by the population during a small time interval $[t,t+\Delta t]$, the population activity is defined as
\begin{equation}
  \label{eq:popact-def}
  A(t)=\lim_{\Delta t\rightarrow 0}\lim_{N\rightarrow\infty}\frac{n(t,t+\Delta t)}{N\Delta t}=\lim_{N\rightarrow\infty}\frac{1}{N}\sum_{i=1}^Ns_i(t).
\end{equation}
Using the population activity, the total input, Eq.~\eqref{eq:def-filters}, can be rewritten as
\begin{equation}
\label{eq:h_int}
h(t)=h_\text{ext}(t)+J\int_0^\infty \epsilon(s)A(t-s)\d s 
\end{equation}
(Fig.~\ref{fig:1}). 
The population activity is uniquely determined by the density $p(\tau,t)$ of refractory states $\tau_i$, where $p(\tau,t)d\tau$ denotes the fraction of neurons with age between $\tau$ and $\tau +  d\tau$ at time $t$. The normalization of this ``refractory density'' is therefore $\int_0^\infty p(\tau,t)\d \tau =1$.
The master equation that governs the time evolution of $p(\tau,t)$ is called \emph{refractory density equation} (RDE) and reads
\begin{equation}
  \label{eq:refr-dens}
  \partial_t p=-\partial_\tau p-\rho(\tau,h(t))p \equiv \op{L}_\tau(h) p.
\end{equation}
The rate of change of the refractory density $p(\tau,t)$ on the right-hand-side of the equation consists of two contributions: first, the advective flux $-\partial_\tau p(\tau,t)$ corresponding to the deterministic and uniform increase of the age in time between spikes; and second, the loss term $-\rho(\tau,h(t))p(\tau,t)$ corresponding to the firing of neurons with age $\tau$ at a  rate $\rho(\tau,h)$. 
All neurons that fire at time $t$ at some age $\tau>0$ will be ``reborn'' at age $\tau=0$. The ``birth rate'' must be equal to the total rate of firing, which is precisely the 
population activity $A(t)$. This conservation of neurons is expressed by the non-local  boundary condition
\begin{equation}
  \label{eq:popact-integr}
  p(0,t) = A(t)=\int_0^\infty \rho\bigl(\tau,h(t)\bigr)p(\tau,t)\d\tau.
\end{equation}
Thus, the population activity can be interpreted as the mean hazard rate $\rho(\tau,h)$ averaged over the refractory density $p(\tau,t)$. The RDE  \eqref{eq:refr-dens} is a first-order partial differential equation, whose right-hand side defines a linear differential operator $\op{L}_\tau(h)$ that will be important for the eigenfunction expansion below. The RDE is equivalent to an integral equation \cite{GerKis14} which enables a rapid numerical integration scheme of the RDE.

%

\section{Spectral decomposition of the refractory density equation}

From a mathematical point of view, the macroscopic dynamics given by the RDE is infinite dimensional because it is in the form of a partial differential equation or integral equation. In the following, we use the spectral decomposition method, or eigenfunction expansion, to derive a hierarchy of ordinary differential equations. The idea is to truncate this hierarchy at a given order to obtain a low-dimensional dynamical system that determines the population activity $A(t)$. The spectral decomposition method has originally been developed for the Fokker-Planck equation \cite{KniMan96,MatGiu02}. Here we apply this method to the refractory density equation \eqref{eq:refr-dens},~\eqref{eq:popact-integr}, which governs general renewal processes.

\subsection{Spectral decomposition}
\label{sec:spectr-decomp}
We first briefly review the spectral decomposition method. The idea is to expand the refractory density $p(\tau,t)$ into eigenfunctions $\{\phi_n(\tau,h)\}_{n\in\mathbb{Z}}$ of the operator $\op{L}_\tau(h)$ at a fixed, constant parameter $h$, i.e.
\begin{equation}
\label{eq:Loperator2}
\op{L}_\tau(h)\phi_n(\tau,h)=\lambda_n(h)\phi_n(\tau,h) 
\end{equation}
where $\lambda_n(h)$ is the eigenvalue associated with the eigenfunction $\phi_n(\tau,h)$. These eigenfunctions form a bi-orthonormal basis with the set of eigenfunctions $\{\psi_n(\tau,h)\}_{n\in\mathbb{Z}}$ of the adjoint operator $\op{L}_\tau^+(h)$, i.e. $\langle\psi_n|\phi_m\rangle=\delta_{n,m}$, where  $\langle\psi|\phi\rangle:=\int_0^\infty\psi(\tau)\phi(\tau)\,d\tau$ denotes the scalar product and the Kronecker symbol $\delta_{n,m}$ is unity if $n=m$ and zero otherwise. Once the basis functions $\phi_n(\tau,h)$ and $\psi_n(\tau,h)$ as well as the eigenvalues $\lambda_n(h)$ have been constructed for any given constant control parameter $h$ (see next section), we obtain a moving basis $\{\phi_n(\tau,h(t))\}$ for time-dependent parameter $h(t)$. Following the traditional approach \cite{KniMan96,MatGiu02}, we represent the population density as a linear superposition of modes, or moving basis functions, $\{\phi_n(\tau,h(t))\}$:
\begin{equation}
  \label{eq:spec-decomp}
  p(\tau,t)=\sum_{n\in\mathbb{Z}}a_n(t)\phi_n(\tau,h(t)).
\end{equation}
The coefficients $a_n(t)$ parameterize how strongly different modes are expressed at time $t$ and will serve as new dynamical variables in our firing-rate model. The zeroth mode $\phi_0(\tau,h)$, which we set as the eigenfunction of $\op{L}_\tau(h)$ associated with eigenvalue $\lambda_0=0$, i.e. $\partial_t\phi_0=\op{L}_\tau(h) \phi_0=0\cdot \phi_0$, is equivalent  to the stationary density $p_0(\tau)$ in the case of constant $h$, i.e. $\phi_0(\tau,h)=p_0(\tau)$ after appropriate normalization of the eigenfunctions (see below). For $n\ne 0$, the coefficients $a_n(t)$ always appear in complex conjugate pairs, $a_{-n}=a_n^*$, which allows us to focus on modes with positive indices $n$. As shown in \emph{Appendix A}, the dynamics of the complex-valued coefficients for $n=1,2,\dotsc$ are given by the ordinary differential equations
\begin{equation}
  \label{eq:an-dyn}
  \dot{a}_n = \lambda_n(h) a_n + \dot{h} \left\{c_{n0}(h)+\sum_{m=1}^\infty \big[c_{nm}(h)a_m+\hat{c}_{nm}(h)a_m^*\big]\right\},
\end{equation}
where the coupling coefficients $c_{nm}(h) = \big\langle \partial_h \psi_n(\tau,h) | \phi_m(\tau,h)  \big\rangle$ and  $\hat{c}_{nm}(h) \equiv c_{n(-m)}(h)= \big\langle \partial_h \psi_n(\tau,h) | \phi_m^*(\tau,h)  \big\rangle$ couple different modes $a_m$. The initial conditions $a_n(0)$ depend on the initial refractory density $p(\tau,0)$. For example, stationary initial conditions, $p(\tau,0)=\phi_0(\tau,h)$, correspond to the initial conditions $a_n(0)=\delta_{n,0}$ (see Appendix Sec.~\ref{sec:append-a:-deriv}). In contrast, if the system is prepared in the fully synchronized state, in which all neurons had a spike immediately before time $t=0^-$, the initial refractory density is $p(\tau,0)=\delta(\tau)$ because the age of all neurons is reset to zero at time $t=0$. Thus, the initial conditions corresponding to the fully synchronized state are $a_n(0)=1$ for all $n\in\mathbb{Z}$. 

Equation \eqref{eq:an-dyn} shows that, in the time-homogeneous case $\dot h=0$, the modes decouple and the coefficients $a_n(t)$, $n=1,2,\dotsc$, relax independently to zero with respective rates determined by  $\lambda_n$. For the stationary mode to be stable, the real part of the eigenvalues $\lambda_n$, $n\ne 0$, must be negative. In the long-term limit only the stationary mode corresponding to $\lambda_0=0$ and $a_0=1$ survives.  Inserting the expansion Eq.~\eqref{eq:spec-decomp} into the boundary condition Eq.~\eqref{eq:popact-integr}, yields the population activity
\begin{equation}
  \label{eq:popact-an}
  A(t) =F_0(h(t))+2\sum_{n=1}^\infty\text{Re}[F_n(h(t))a_n(t)],
\end{equation}
where $F_n(h):=\phi_n(0,h)$. 
The time evolution of $A(t)$ is then completely determined by the dynamics Eq.~\eqref{eq:an-dyn} of the modes $a_n$.

\subsection{Eigenvalues and eigenfunctions of the refractory density equation}
\label{sec:eigenv-eigenf-refr}

The eigenfunctions $\phi_n(\tau,h)$ solve the eigenvalue problem Eq.~\eqref{eq:Loperator2} and
respect the boundary condition, Eq.~\eqref{eq:popact-integr}, that is
\begin{align}
  \partial_\tau \phi_n(\tau,h) &= - \rho(\tau,h) \phi_n - \lambda_n \phi_n, \label{eq:phi_n-dynamics} \\
  \phi_n(0,h) &\equiv F_n(h)= \int_0^\infty\rho(\tau,h)\phi_n(\tau,h)\d\tau. \label{eq:boundarycondition2}
\end{align}
Using the definition Eq.~\eqref{eq:survprob} of the survival probability $S(\tau,h)$, the solution of Eq.~\eqref{eq:phi_n-dynamics} reads
\begin{equation}
\label{eq:phin_general}
\phi_n(\tau,h)=F_n(h)S(\tau,h)\mathrm{e}^{-\lambda_n\tau}
\end{equation}
Substituting Eq.~\eqref{eq:phin_general}
into the boundary condition Eq.~\eqref{eq:boundarycondition2} then yields
\begin{equation}
\label{eq:condition2}
P_L(\lambda_n,h)=1,
\end{equation}
where $P_L(s,h)=\int_0^\infty e^{-s\tau}P(\tau,h)\d\tau$ denotes the Laplace transform of the ISI density Eq.~\eqref{eq:ISI_dens}. Equation \eqref{eq:condition2} is the characteristic equation for the eigenvalues $\lambda_n$. As we show in the Appendix, Sec.~\ref{sec:alt-deriv}, the characteristic equation can also be derived directly from the refractory density equation in Laplace space. To find non-trivial complex solutions of Eq.~\eqref{eq:condition2}, $P_L(\lambda,h)$ is extended to the complex plane by analytic continuation.


An alternative representation of the characteristic equation for the eigenvalues $\lambda_n$ can be obtained in terms of the survival probability $S(\tau,h)$.
Using the relation $P(\tau,h)=-\partial_\tau S(\tau,h)$ and integrating by parts, we find that the non-trivial eigenvalues $\lambda_{n \neq 0}$ are given by the complex roots of the Laplace transform of the survival probability,
\begin{equation}
\label{eq:condition3}
S_L(\lambda_n,h)=0,
\end{equation}
where $S_L(s,h)=\int_0^\infty e^{-s\tau}S(\tau,h)\,d\tau$.
We can conclude from Eq.~\eqref{eq:condition3} that if $\lambda_n$ is an eigenvalue, also its complex conjugate $\lambda_n^*$ is an eigenvalue, with corresponding eigenfunction $\phi_n^*$.
In the following, we will use the definitions $\lambda_{\shortminus n}=\lambda_n^*$ and $\phi_{\shortminus n}=\phi_n^*$, so that the sums over the spectrum of the evolution operator range over all integer numbers.

As $\op{L}_\tau$ is not Hermitian, we also need to define the set of eigenfunctions $\{\psi_n\}$ of the adjoint operator $\op{L}_\tau^{+}$.
Using the properties of the scalar product introduced above, we have
\begin{align}
  \langle\op{L}_\tau^+\psi|\phi\rangle&=	\langle\psi|\op{L}_\tau\phi\rangle= \int_{0}^{\infty}\psi(\tau)[-\partial_{\tau}-\rho(\tau)]\phi(\tau)d\tau  \nonumber \\
                                      &= \psi(0)\phi(0)+ \int_{0}^{\infty}\phi(\tau)[\partial_{\tau}-\rho(\tau)]\psi(\tau)d\tau  \nonumber \\
	&= \int_{0}^{\infty}\big\{\psi(0)\rho(\tau)+[\partial_{\tau}-\rho(\tau)]\psi(\tau)\big\}\phi(\tau)d\tau,  \nonumber
\end{align}
from which we find the explicit form
$$
\op{L}_\tau^+\psi(\tau)=[\partial_{\tau}-\rho(\tau)]\psi(\tau)+\rho(\tau)\psi(0).
$$
The adjoint operator $\op{L}_\tau^+$ has the same eigenvalues as $\op{L}_\tau$ and its eigenfunctions $\{\psi_m\}$ then obey the dynamics
\begin{equation}\label{eq:psi_m-dyn}
\partial_\tau \psi_m(\tau) = \big[ \lambda_m + \rho(\tau) \big] \psi_m + \rho(\tau) \psi_m(0).
\end{equation}
Without loss of generality, we are free to choose $\psi_m(0)=1$ and fix the remaining factor $F_n(h)$ in Eq.~\eqref{eq:phin_general} by imposing the bi-orthonormality relation $\langle\psi_m|\phi_n\rangle=\delta_{n,m}$. 
The solution of Eq.~\eqref{eq:psi_m-dyn} is thus given by
\begin{eqnarray}
  \label{eq:psin}
  \psi_m(\tau)&=&\exp\left(\lambda_m \tau + \int_0^\tau \rho(s)ds \right)\times \nonumber\\
  &&\left[ 1 - \int_0^\tau \rho(x) \exp\left( -\lambda_m x - \int_0^x \rho(s) ds\right)dx \right] \nonumber\\
  &=&\frac{\mathrm{e}^{\lambda_m  \tau}}{S(\tau)} \left( 1-\int^\tau_0 P(x)\mathrm{e}^{-\lambda_m x}\,\d x \right)\ ,
\end{eqnarray}
where we used the definition Eq.~\eqref{eq:survprob} of the survival probability $S(\tau)$.
In particular, we have that $\psi_0(\tau) \equiv 1$ and, consequently, also $a_0(t) \equiv 1$ because $a_0 = \langle \psi_0 | p\rangle = \int_0^\infty p(\tau,t) d\tau = 1$ due to the normalization of the refractory density $p(\tau,t)$. 

Using the solutions of $\phi_n$ and $\psi_m$, Eqs.~(\ref{eq:phin_general}) and (\ref{eq:psin}), and equating their scalar product to unity, we find that
\begin{equation}
  \label{eq:phi0-lap}
F_n(h)=  -\frac{1}{P_L'(\lambda_n,h)},
\end{equation}
where the prime denotes the derivative with respect to the first argument.
Because $\lambda_0=0$ and $P_L(s,h)$ is the moment generating function, the factor $F_0(h) = -1/P_L'(0,h)$ coincides with the mean rate $r=1/\langle\tau\rangle$, where $\langle\tau\rangle = \int_0^\infty \tau P(\tau,h) d\tau = -P_L'(0,h)$ is the mean ISI. 
In other words, $F_0(h)$ is the stationary transfer function (``f-I curve'').

In conclusion, for a given momentary input $h$, we have found that the eigenvalues $\lambda_n(h)$ are given as the complex solutions of the characteristic equations
\begin{equation}\label{eq:eigenvalue-formula}
P_L \big(\lambda, h\big) = 1 \quad \text{or}\quad S_L\big(\lambda, h\big) = 0 \ ,
\end{equation}
where $P_L$ and $S_L$ denote the Laplace transforms of the ISI density and survival probability, respectively.
Moreover, the eigenfunctions are given by
\begin{subequations}\label{eq:phin_psim}
\begin{align}
\phi_n(\tau,h) &= -\frac{S(\tau,h)\mathrm{e}^{-\lambda_n(h) \tau}}{P_L'\big(\lambda_n(h), h\big)}  \\
\psi_n(\tau,h) &= \frac{\mathrm{e}^{\lambda_n(h) \tau}}{S(\tau,h)}  \left[ 1 - \int_0^\tau P(s,h) \mathrm{e}^{-\lambda_n(h) s} \d s \right],
\end{align}
\end{subequations}
for all $n\in\mathbb{Z}$. Note that the macroscopic population dynamics, Eqs.~\eqref{eq:an-dyn} and \eqref{eq:popact-an}, is fully expressed in terms of the hazard rate, the survival probability or the ISI density through the equations (\ref{eq:eigenvalue-formula}) and (\ref{eq:phin_psim}). Thus, these equations link the dynamics at the macroscopic scale with the single neuron dynamics at the microscopic scale.

%

%
%
%
%

\subsection{Low-dimensional firing rate model}
\label{sec:low-dimens-firing}
The purpose of the eigenfunction expansion of the refractory density operator $\op{L}_\tau$ is that it allows us to systematically reduce the complexity of the infinite-dimensional evolution equation Eq.~\eqref{eq:refr-dens}.
So far, the infinite system of ordinary differential equations, Eq.~\eqref{eq:an-dyn}, obtained from the eigenfunction expansion is still infinite-dimensional. However, assuming a sufficiently large spectral gap between the first $m$ nontrivial eigenvalue and the rest of the eigenvalue spectrum, i.e.  $\mathrm{Re}(\lambda_n) \ll \mathrm{Re}(\lambda_m) < 0$ for $n > m$, higher-order modes $a_n$ will decay much faster than the first $m$ modes. The rapid decay of higher-order modes enables us to truncate the expansion, Eq.~\eqref{eq:spec-decomp}, after the term $n=m$, or equivalently,  to set $a_n\equiv 0$ for $n>m$. To obtain a low-dimensional firing rate model, we also assume that the control parameter $h(t)$ follows low dimensional dynamics, e.g., the first-order dynamics 
\begin{equation}
  \label{eq:h-first-ord-FR}
  \tau_h\od{h}{t}=-h+I_\text{ext}(t)+JA(t),
\end{equation}
corresponding to exponential filter functions $\kappa(t)=\epsilon(t)=\tau_h^{-1}e^{-t/\tau_h}\Theta(t)$ in Eq.~\eqref{eq:def-filters}. 

The truncation of the eigenfunction expansion Eq.~\eqref{eq:spec-decomp} at different orders yields a hierarchy of firing rate models with increasing dimensionality. The simplest case is a truncation at the stationary mode $n=0$, yielding the \emph{zeroth-order approximation}
\begin{equation}
  \label{eq:zeroth-order}
  A(t)\approx F_0(h(t)),
\end{equation}
where $F_0\big(h(t)\big)$ is the stationary f-I curve, or transfer function, of the neurons.
The approximation Eq.~\eqref{eq:zeroth-order} together with Eq.~\eqref{eq:h-first-ord-FR} recovers the classical one-dimensional firing rate model. Because it only accounts for the stationary mode $\phi_0(\tau,h)$, the zeroth-order approximation cannot capture dynamics beyond that of $h(t)$ (Eq.~\eqref{eq:h-first-ord-FR}); in particular, it cannot capture dynamical effects of refractoriness.

At the first nontrivial order, we retain the dynamics of $a_1(t)$ but neglect all higher-order modes $a_n$ for $n\ge 2$, which yields the \emph{first-order approximation}
\begin{subequations}\label{eq:first-mode-approx}
\begin{align}
A(t) &= F_0\big(h\big) -  2\, \mathrm{Re} \big[ a_1(t) / P_L' \big(\lambda_1(h)\big) \big] \label{eq:fma_a}\\
\od{a_1}{t} &=  \lambda_1(h) a_1 + \od{h}{t} \big[ c_{10}(h) + c_{11}(h)a_1 + c_{1(-1)}(h)a_1^* \big].   \label{eq:fma_b}
\end{align}
\end{subequations}
The initial condition is $a_1(0)=0$ for the stationary initial state and $a_1(0)=1$ for the synchronized initial state (cf. Sec.~\ref{sec:spectr-decomp}). The steady-state transfer function $F_0(h)=-1/P_L'(0,h)$ represents the skeleton of the dynamics, around which contributions of the first mode $a_1(t)$ describes the relaxation dynamics towards $F_0(h)$. Note that the dynamics of $a_1$ is two-dimensional because $a_1$ is complex-valued. In total, the firing rate model Eq.~\eqref{eq:first-mode-approx} together with the dynamics of $h$, Eq.~\eqref{eq:h_int}, defines a $(2+d)$-dimensional dynamical system, where $d$ is the dimensionality of the dynamics of $h(t)$ ($d=1$ for $h$ given by Eq.~\eqref{eq:h-first-ord-FR}). As we will show, the three-dimensional firing rate model approximates the full infinite-dimensional population activity Eqs.~(\ref{eq:refr-dens}) and (\ref{eq:popact-integr}) to great extent. 

In general, \emph{higher-order approximations} can be obtained by neglecting all amplitudes $a_n$ for some integer $n>m$ in Eqs.~\eqref{eq:an-dyn} and \eqref{eq:popact-an} giving rise to a $(2m+d)$-dimensional firing rate dynamics.


\subsection{Linearized firing-rate dynamics}
\label{sec:linearizeddyn}

An important characteristic of the population dynamics is the linear response to weak time-dependent modulation of the control parameter $h$: $h(t)=h_0+h_1(t)$. For sufficiently weak  modulation $h_1(t)$, the firing rate dynamics can be linearized about the steady-state population activity $A_0=F_0(h_0)$. For the first-order approximation, Eq.~\eqref{eq:first-mode-approx}, the linearized dynamics reads
\begin{subequations}\label{eq:first-mode-approx-lin}
\begin{align}
A(t) &= F_0\big(h_0\big)+F_0'(h_0)h_1(t) +  2\, \mathrm{Re} \big[ F_1(h_0)a_1(t)\big] \label{eq:fma_a-lin}\\
\od{a_1}{t} &=  \lambda_1(h_0) a_1 + \od{h_1}{t}c_{10}(h_0)\label{eq:fma_b-lin}
\end{align}
\end{subequations}
(see Appendix, Sec.~\ref{sec:linearizeddyn-eig-append}). Applying the Fourier transform, $\tilde{f}(\omega)=\int_{-\infty}^\infty e^{-i\omega t}f(t)\,dt$, to Eq.~\eqref{eq:first-mode-approx-lin}, the population activity can be related to the input in frequency space by 
\begin{equation}
\tilde{A}_1(\omega) = \tilde\chi_{1,h}(\omega) \tilde{h}_1(\omega)
\end{equation}
where we introduced the linear response function, or susceptibility,
\begin{equation}
  \label{eq:lin-respon}
  \tilde\chi_{1,h}(\omega)=F_0'(h_0)  + i\omega \left( \frac{F_1(h_0) c_{10}(h_0)}{i\omega -  \lambda_1(h_0)} +\frac{F_1^*(h_0) c^*_{10}(h_0)}{i\omega - \lambda^*_1(h_0)}\right).
\end{equation}
Corresponding formulas that account for higher-order approximations are given in the Appendix, Sec.~\ref{sec:linearizeddyn-eig-append}. The linear response function, Eq.~\eqref{eq:lin-respon}, will be used in Sec.~\ref{sec:popul-dynam-resp}, to compare the response properties of the first-order approximation and the full refractory density dynamics for arbitrary weak stimuli.


\section{Spectral decomposition for specific neuron models}
\label{sec:spec_decomp_specific}

To illustrate the reduction to low-dimensional firing-rate dynamics, we now apply the theory to several examples.

\begin{figure*}[!htbp]
\begin{center}
\includegraphics[scale=1]{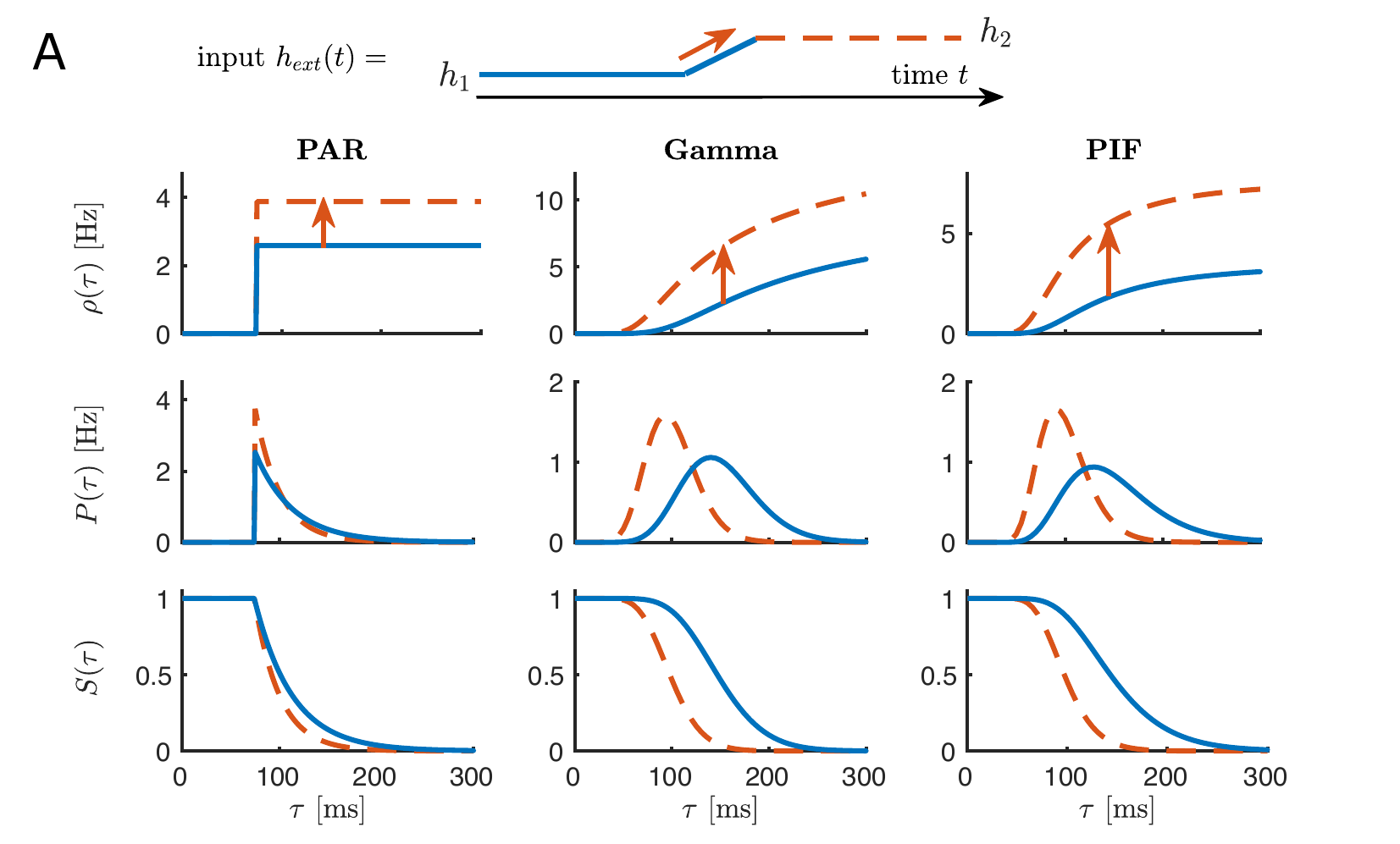}\\
\includegraphics[scale=1]{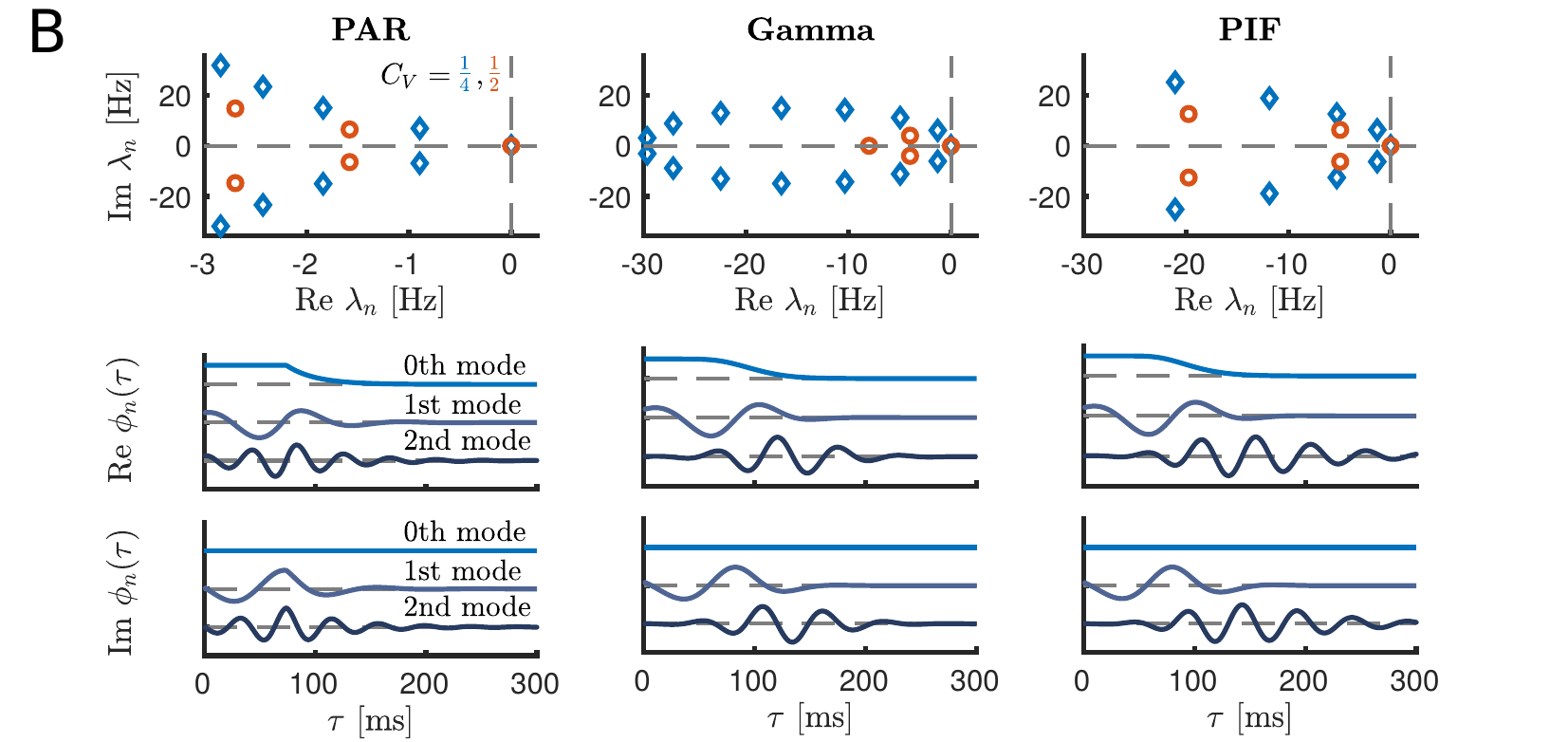}
\caption{A: Hazard rate $\rho(\tau)$, ISI density $P(\tau)$ and survival probability $S(\tau)$ for three model neurons with constant input $h$ that is increased from $h(t)=h_1 $ (blue) to $h(t)=h_2$ (red) and a CV of $\cv=1/\sqrt{15}$: Poisson neuron with absolute refractoriness (PAR, left), Gamma neuron model (Gamma, middle), perfect integrate-and-fire neuron (PIF, right).
B: Eigenvalues (top) for the three model neurons without input current and a CV of $\cv=1/\sqrt{15}$ (blue diamonds) and $\cv=1/2$ (red circles), respectively.
Real and imaginary parts (normalized) of the first three eigenmodes $\phi_{0,1,2}(\tau)$ (bottom) with constant input $h$ and $\cv=1/\sqrt{15}\approx 1/4$.}
\label{fig:2}
\end{center}
\end{figure*}

\subsection{Poisson model with absolute refractoriness (PAR)}
\label{sec:poisson-model-with}

First, we consider neurons modeled as an inhomogeneous Poisson process with absolute refractory period (PAR), i.e. a neuron fires with a rate $\nu\big(h(t)\big)$ if it is not in its absolute refractory state: After firing, a neuron is inactive during an absolute refractory period of length $\Delta$.
The hazard rate can thus be written as:
\begin{equation}
\label{eq:Pois_hazard}
\rho(\tau,h)=\nu(h)\Theta(\tau-\Delta)
\end{equation}
with $\Theta(s)$ the Heaviside step function.
As an example used in our figures, we choose an exponential rate function $\nu(h)=\nu_0\exp[(h-\theta)/\delta]$ with constants $\nu_0, \theta$ and $\delta$ that can be interpreted as the instantaneous rate $\nu_0$ at threshold, threshold $\theta$ and threshold softness $\delta$ (modeling the noise strength).
For a given absolute refractory period $\Delta$, 
the mean firing rate is $r=\nu_0/(1+\Delta \nu_0)$ and the coefficient of variation (CV) is $\cv=1/(1+\Delta\nu_0)$.
In Fig.~\ref{fig:1} (B, left column), we plot the hazard rate $\rho(\tau,h)$, survival probability $S(\tau,h)$ and ISI density $P(\tau,h)$ for two constant input levels $h=h_{1,2}$, such that for $h=h_1$ the neuron has unit mean rate $r=1$ and $\cv=1/\sqrt{15}$. 
An increase in input $h_1 \nearrow h_2$ leads to a parallel shift from $\nu(h_1)$ to $\nu(h_2)$.
The ISI density $P(\tau,h)$ describes an exponential decay starting from $P(\Delta,h)= \nu(h)$ with rate $\nu(h)$, so that an increase in input results in a higher peak at the refractory period $\tau=\Delta$ and a faster decay for $\tau>\Delta$.
This faster decay for larger input $h_2>h_1$ is also reflected by the survival probability $S(\tau,h)$.

The Laplace transform of the ISI density reads
$$
P_L(\lambda) = \frac{\nu}{\nu+\lambda} \exp(-\lambda	\Delta).$$
Using the characteristic equation, Eq.~\eqref{eq:eigenvalue-formula}, we find the eigenvalues
\begin{equation}\label{eq:Pois_lambdan}
\lambda_n=\frac{1}{\Delta}W_n(\Delta \nu e^{\nu\Delta}) - \nu,
\end{equation}
where $W_n(z)$ is the $n$th branch of the Lambert $W$-function.
We can explicitly express the eigenfunctions $\phi_n$ of the operator $\op{L}_\tau$ and $\psi_n$ of the adjoint operator $\op{L}_\tau^+$
as
\begin{align}
\phi_n(\tau) &=\frac{\nu+\lambda_n}{1+\Delta(\nu+\lambda_n)}e^{-\lambda_n\tau}\times\nonumber\\
&\hspace{.5cm} \times\left[\Theta(\Delta-\tau) +  \Theta(\tau-\Delta)e^{-\nu(\tau-\Delta)}\right],\\
\psi_n(\tau)&=\Theta(\Delta-\tau)e^{\lambda_n\tau} +  \Theta(\tau-\Delta)e^{\lambda_n\Delta}.
\end{align}
The normalization constants thus read
\begin{equation}
  \label{eq:Fn-par}
F_n(h) = \frac{\nu(h) + \lambda_n(h)}{1+ \Delta \big( \nu(h) + \lambda_n(h) \big)}
\end{equation}
and the transfer function is
\begin{equation}
  \label{eq:F0-par}
F_0(h) = \frac{\nu(h)}{1+ \Delta \nu(h)},
\end{equation}
where we used that $\lambda_0 = 0$.
In Fig.~\ref{fig:2} (left column), we plot the first eigenvalues $\lambda_n$ (top panel) together with
the first three eigenfunctions $\phi_{0,1,2}(\tau)$ (lower panels) for constant input $h=h_0$, unit mean rate $r=1$ and CV as indicated in the caption.
The spectrum of the PAR model lies in the left complex half-plane with one null eigenvalue $\lambda_0=0$.
The other eigenvalues have negative real part and come in complex conjugate pairs.
Not only does a larger CV increase the absolute value of the real parts of the eigenvalues, but also the spectral gap between the first and the second pairs of complex conjugate eigenvalues increases for larger CV.
As to the eigenfunctions, note that the zeroth eigenfunction coincides with the survival probability, so that  $\mathrm{Re}\,\phi_0(\tau)= S(\tau)$ and $\mathrm{Im}\,\phi_0(\tau)=0$.
Higher modes resemble wavelets that successively increase both in width and frequency.

The coupling coefficients $c_{nm}= \langle \partial_h \psi_n | \phi_m  \rangle$ are obtained by differentiating $\psi_n$ with respect to $\nu$ using $\partial_h \psi_n\big( \nu(h) \big)= \partial_\nu \psi_n(\nu) \nu'(h)$, which yields
\begin{align}
	c_{nn}=&\frac{\lambda_n\Delta(1+\frac{1}{2}\Delta(\nu+\lambda_n))}{\nu(1+\Delta(\nu+\lambda_n))^2}\nu'(h)\\
        c_{nm}=&\frac{\lambda_n(\nu+\lambda_m)}{\nu(\lambda_n-\lambda_m)(\nu+\lambda_n)(1+\Delta(\nu+\lambda_m))}\nu'(h).\label{eq:Pois_cnm}
\end{align}
The prime denotes differentiation with respect to $h$.

\subsection{Gamma model}
\label{sec:gamma-model}

As another simple renewal model, we consider the gamma process, whose ISI distribution is given by the gamma distribution with integer shape parameter (also called Erlang distribution) and an input-dependent rate parameter
\begin{equation}
\label{eq:Gam_dist}
P(\tau)=\frac{\nu^\alpha}{(\alpha-1)!}\tau^{\alpha-1}e^{-\nu\tau}.
\end{equation}
Here, $\nu(h)\in\mathbb{R}^+$ and $\alpha\in\mathbb{Z}^+$ are rate and shape parameters, respectively. The gamma distribution has frequently been used to model ISI densities of neurons \citep{Ost11}.
For integer $\alpha$, the renewal process can be generated by a neuron model that exhibits $\alpha$ discrete states $S_1,\dots,S_\alpha$. Transitions from state $S_k$ to $S_{k+1}$ occur at an input-dependent rate $\nu(h)$, and $S_{\alpha+1}$ is identified with $S_1$. Only the transition from $S_\alpha\rightarrow S_1$ corresponds to a spiking event. The total firing rate is thus $r=F_0(h)=\nu(h)/\alpha$ and the CV is $\cv=1/\sqrt{\alpha}$. For $\alpha=1$, we retrieve the Poisson process with $\cv=1$.
If $\alpha\ge 2$, the neuron has to pass through $\alpha-1$ intermediate states before it can eventually fire, and the resulting spike train exhibits refractoriness. In this case, the Gamma process is more regular than a Poisson process.

The ISI density, hazard rate and survival probability of a gamma neuron model with constant rate is illustrated in Fig.~\ref{fig:1}B, middle.
In contrast to the PAR model, an increase in input $h_1 \nearrow h_2$ not only leads to an upwards shift of the hazard rate $\rho(\tau)$, but also results in an earlier offset, thereby decreasing the (effective) refractory time. 
At the same time, the ISI density $P(\tau)$ moves to the left and becomes narrower for larger input $h_2>h_1$.
This narrowing effect goes along with a faster decay  of the ISI density with dominant rate $\nu(h)$, which can also be seen in the survival probability $S(\tau)$.

For the ISI density Eq.~\eqref{eq:Gam_dist} of the Gamma process, the Laplace transform is
\begin{equation}
P_L(\lambda)=\Big(\frac{\nu}{\nu+\lambda}\Big)^\alpha,
\end{equation}
where we have omitted the explicit dependence on $h$.
Condition Eq.~\eqref{eq:eigenvalue-formula} leads to the (discrete and finite) set of eigenvalues
\begin{equation}
\label{eq:gammaeig}
\lambda_n=\nu\left(\exp(2\pi i n/\alpha)-1\right), \quad n=0,..., \alpha-1.
\end{equation}
We can also express the eigenvalues in terms of the rate $r$ and CV as
\begin{equation}
\lambda_n=rC_V^{-2}\left(\exp(2\pi iC_V^2 n)-1\right).
\end{equation}
The normalization constants $F_n(h)$ simplify as
\begin{equation}
F_n(h)=\frac{\nu(h) + \lambda_n(h)}{\alpha},
\end{equation}
which allows us to compute the eigenfunctions $\phi_n$ and $\psi_m$.
As of yet, the integrals in the coupling coefficients $c_{nm}$ elude an analytic solution.
In Fig.~\ref{fig:2} (middle column), we illustrate the first eigenvalues $\lambda_n$ (top) as well as the eigenfunctions $\phi_{0,1,2}(\tau)$ (bottom) for unit rate $r=1$ and $\cv=1/\sqrt{\alpha}$ with $\alpha=15$.
Remarkably, for integer $\alpha$, there is only a finite number of eigenvalues $\lambda_0, \dots, \lambda_{\alpha-1}$ and they describe a circle with one null eigenvalue $\lambda_0=0$.
As in the PAR model, the other eigenvalues come in complex conjugate pairs and the spectral gap increases for larger CV, that is, for smaller $\alpha$.
Note that while the first non-trivial eigenvalue of the Gamma model lies in the same range as that of the PAR model, the gaps between the other eigenvalues are distinctly larger in the Gamma model than in the PAR model, cf.~the different scales of the real-axis in Fig.~\ref{fig:2}.

\subsection{Integrate-and-fire models driven by Gaussian white noise}
\label{subsec:if-models}

Integrate-and-fire (IF) neurons driven by white Gaussian noise constitute an important modeling tool in computational neuroscience \cite{Bur06a} and belong to the class of renewal processes. Their population activity can thus be analyzed within the framework of the refractory density approach. The model is defined by  the subthreshold membrane potential $v$ that obeys the Langevin equation combined with an algorithmic fire-and-reset rule
\begin{gather}\label{eq:ifdyn_general1}
  \dot{v} = f_\text{model}(v) + \mu(h) + \sqrt{2D}\, \xi(t),\\
  \text{if } v=\vth, \text{ then } v\rightarrow \vreset \nonumber.
\end{gather}
According to the fire-and-reset rule, the neuron elicits a spike once $v(t)$ crosses a threshold value $\vth$ and is subsequently reset to a reset value $\vreset$. In Eq.~\eqref{eq:ifdyn_general1},  time is measured in units of the membrane time constant $\tau_m$. The parameters $\mu$ and $D$ represent the mean input and noise intensity, respectively. For simplicity, we here assume that only $\mu=\mu(h)$ depends on the input $h$ as defined in Eq.~\eqref{eq:h_int}. 
In general, however, both $\mu$ and $D$ can depend on (distinct) input parameters $h_1, h_2, \dotsc$ as would be the case in a diffusion approximation of inhomogeneous Poisson inputs \cite{BruHak99,Bur06b,MatGiu02}. Furthermore,
$\xi(t)$ is a zero-mean Gaussian white noise with $\langle \xi(t)\xi(t')\rangle=\delta(t-t')$. The two simplest, yet important choices for the model-specific function $f_\text{model}$ are $f_\text{PIF} (v) = 0$ and $f_\text{LIF} (v) = -v$
for the perfect integrate-and-fire (PIF) and the leaky integrate-and-fire (LIF) neuron model, respectively.

The ISI density $P(t)$ of an IF model driven by white noise is equal to the first-passage-time density $\hat{P}(t,v_0)$ for the time $t>0$ of the first threshold crossing starting with an initial value $v(0)=v_0$ at the reset potential $v_0=\vreset$: $P(t)=\hat{P}(t,v_0)|_{v_0=\vreset}$. Similarly, let us write the survival function as $S(t)=S(t,v_0)|_{v_0=\vreset}$ to explicitly account for the initial value. Using a similar notation for the Laplace transforms of the ISI density and the survival functions, we write $P_L(s)=\hat{P}_L(v_0;s)|_{v_0=\vreset}$ and $S_L(s)=\hat{S}_L(v_0;s)|_{v_0=\vreset}$, respectively, where the Laplace argument $s$ is considered as a parameter. For any fixed parameter $s$, the Laplace transforms are given by the solutions of the following boundary value problems \cite{Tuc88}:
\begin{subequations}\label{eq:fpt_ode}
\begin{gather}
s\hat{P}_L - (f_\text{model}(v_0) +\mu) \hat{P}_L' - D \hat{P}_L''= 0, \\
      \hat{P}_L'(a;s)=0,\qquad
  \hat{P}_L(\vth;s) = 1,
\end{gather}
\end{subequations}
and \cite{Gar85,Tuc88,ErmTer10}
\begin{subequations}\label{eq:sfunc_ode}
\begin{gather}
s\hat{S}_L - (f_\text{model}(v_0) +\mu) \hat{S}_L' - D \hat{S}_L'' = 1,\\
\hat{S}_L'(a;s) =0 ,\qquad \hat{S}_L(\vth;s)=0.
\end{gather}
\end{subequations}
Here, $a\le\vreset$ is a left reflecting boundary that can be set to $a=-\infty$ for standard IF models. The boundary value problems Eqs.~\eqref{eq:fpt_ode} and \eqref{eq:sfunc_ode} can be solved analytically for simple cases such as the perfect or leaky IF models, see below. For general nonlinear IF models, these equations have to be integrated numerically. In this case, instead of the boundary condition at $a=-\infty$, it is convenient to use a lower reflecting boundary at a sufficiently low but finite value $a<\vreset$.

\subsubsection{Perfect integrate-and-fire model}

\begin{figure*}[t]
\begin{center}
\includegraphics[scale=1]{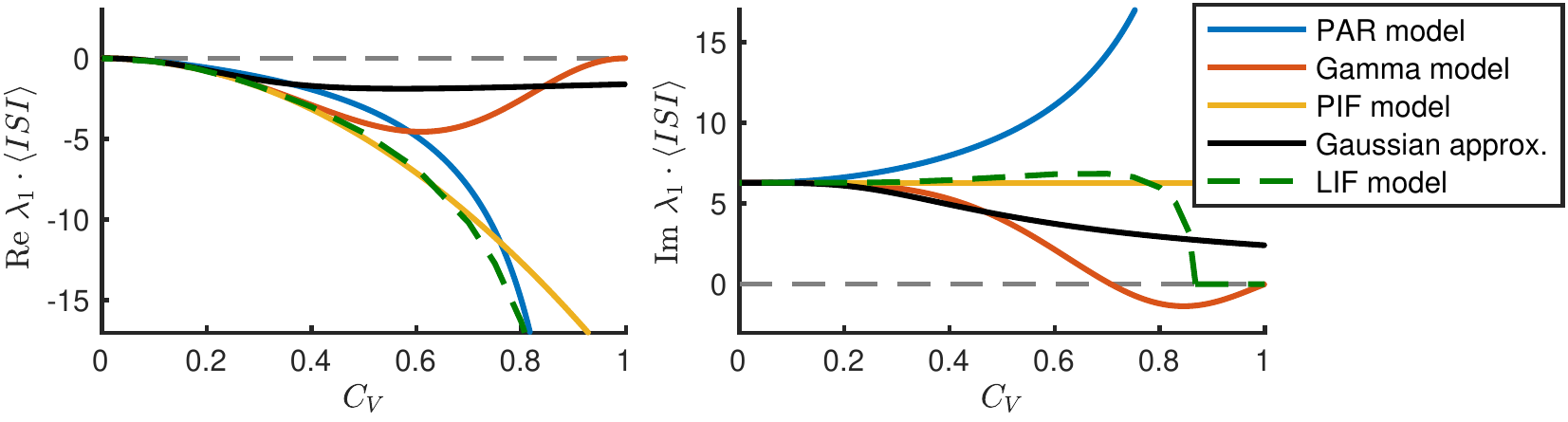}
\caption{Dominant eigenvalue differs for different renewal models with the same mean and variance of the interspike intervals. Gaussian approximation of the first eigenvalue (black) presents a valid fit of the analytically determined eigenvalues of different renewal processes (see legend) for low CV.
Shown are the real (left panel) and imaginary (right) parts of the first eigenvalue divided by the mean rate $r=1/\langle ISI \rangle$.
}
\label{fig:6}
\end{center}
\end{figure*}

The perfect integrate-and fire (PIF) model is obtained by setting $f_\text{model}(v) = 0$ in Eq.~\eqref{eq:ifdyn_general1}. The PIF model \cite{GerMan64} and extensions thereof  \cite{FisSch12,SchMik13,SchDro15} are canonical models for tonically spiking neurons in the mean-driven regime  and have successfully been used to explain stationary ISI statistics of various cell types.
The membrane potential $v$ in the PIF model can be seen as an overdamped Brownian motion with drift $\mu(h)$.
Without loss of generality, we set $\vreset=0$. 
The ISI density is an inverse Gaussian distribution (Appendix, Sec.~\ref{sec:known-stat-perf}), for which the Laplace transform is well-known \cite{Hol76,BulLow94,ChaLon05,Vre10}:
\begin{equation}
  \label{eq:PL-pif}
P_L(s)=\exp\lreckig{\frac{1}{C_V^2}\left(1-\sqrt{1+\frac{2C_V^2s}{r}}\right)}.
\end{equation}
Here, $r=\mu/\vth$ and $C_V^2=2D/(\mu\vth)$ are the rate and coefficient of variation of the ISIs, respectively. 
Solving the characteristic equation \eqref{eq:condition2}, $P_L(\lambda)=1$ yields the eigenvalues
\begin{equation}\label{eq:PIF_eigenvalues_rateCV}
\lambda_n=- 2\pi^2rC_V^2n^2+2\pi r i\:n, \quad n=0,1,2,\dots.
\end{equation}
The normalization constants can be found as
\begin{equation}
\label{eq:PIFnorm}
F_n(h) =\sqrt{r^2+2rC_V^2\lambda_n}
\end{equation}
which we can use to plot the eigenfunctions $\phi_{n}(\tau)$ (bottom) in Fig.~\ref{fig:2} (right column). 
There, we also plot the first eigenvalues $\lambda_n$ (top) for unit mean rate $r=1$ and $\cv=1/\sqrt{15}$.
As for the PAR and the Gamma model, there is one trivial eigenvalue $\lambda_0=0$ and the others have negative real part and come in complex conjugate pairs.
Again, the first non-trivial eigenvalue has similar real and imaginary parts as in the other two models.
The spread between eigenvalues is, however, more accentuated for the PIF model and the spectral gap between the first and second eigenvalue is striking, especially when increasing the CV.

\subsubsection{Leaky integrate-and-fire model}

\begin{figure*}[t]
\begin{center}
\includegraphics[scale=1]{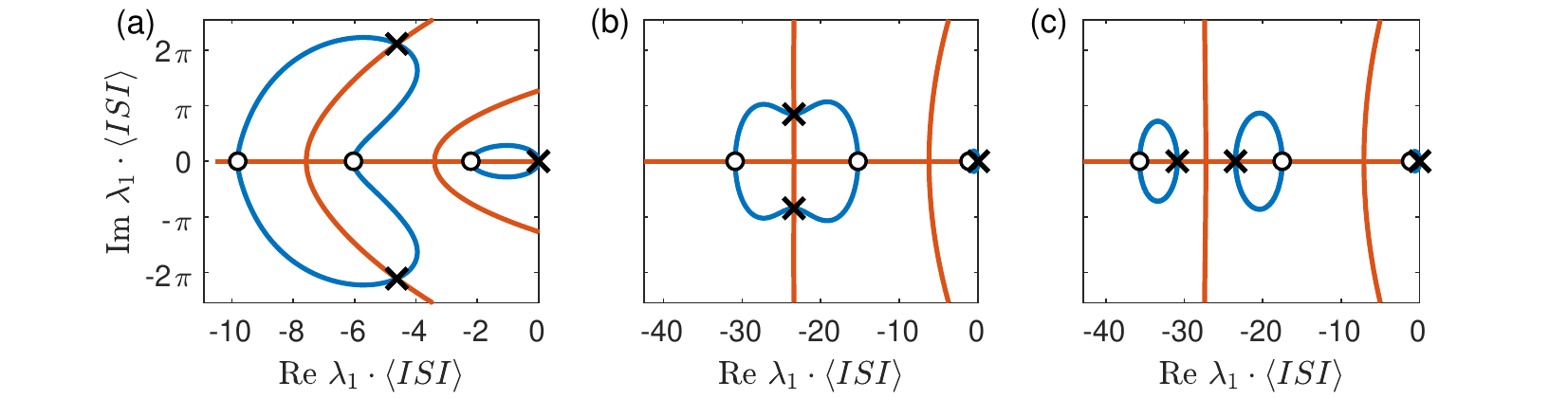}
\caption{First eigenvalue $\lambda_1$ and its complex conjugate $\lambda_1^*$ of the LIF model can be found numerically as the intersection points (black crosses) of the contour lines $\{\mathrm{Re}\,P_L(s)=1\}$ (blue) and $\{ \mathrm{Im}\, P_L(s)=0 \}$ (red) for increasing CV: (a) $C_V = 0.5$,  (b) $C_V = 0.86$ and (c) $C_V = 0.88$.
The white dots represent singularities, which can be identified when inspecting $S_L(x)$ along the real axis.
The banana shape of the blue contour line becomes more rectified for larger CV, implicating a gradual increase in the absolute real value $|\mathrm{Re}\,\lambda_1|$ while the imaginary part remains constant at around $\mathrm{Im}\,\lambda_1 = 2\pi$.
Close to the transition from the mean- to noise-dominated regime, a dip at the extrema of the blue contour line emerges and the pair of complex conjugate eigenvalues moves towards the real axis with roughly constant real value.
Eventually, the real contour splits into two and crosses the real axis in parallel to and on each side of the imaginary (non-trivial) contour line, giving rise to a pair of purely real eigenvalues $\lambda_1 < \lambda_2$.
}
\label{fig:LIFcontour}
\end{center}
\end{figure*}

The leaky integrate-and-fire (LIF) model is a widely used neuron model in theoretical neuroscience \cite{DayAbb05,GerKis02}.
The specific model function in Eq.~\eqref{eq:ifdyn_general1} now reads $f_\text{LIF}(v) = -v$ and we consider the same fire-and-reset rule as before.
 Without loss of generality, we can set $\vreset=0$ and $\vth=1$.
The Laplace transform of the ISI density is known in terms of special functions \cite{LinLSG02}
\begin{equation}
P_L(\lambda)  = \mathrm{e}^\delta \, \frac{\mathcal{D}_{-\lambda} \Big( \frac{\mu}{\sqrt{D}}\Big)}{\mathcal{D}_{-\lambda} \Big( \frac{\mu-1}{\sqrt{D}}\Big)}, \quad \delta = \frac{1+2\mu}{4D},
\end{equation}
where $\mathcal{D}_\alpha(z)$ denotes the parabolic cylinder function.
To find the eigenvalues $\lambda_n$ satisfying the eigenvalue equation $P_L(\lambda_n)=1$, it is instructive to exploit the contour lines $\big\lbrace \mathrm{Re}\, P_L(\lambda) = 1\big\rbrace $ and $\big\lbrace \mathrm{Im}\, P_L(\lambda) = 0\big\rbrace $ in the complex plane, see Fig.~\ref{fig:LIFcontour}.
The eigenvalues are the intersection points of the contour lines.
Note that some of the intersections in Fig.~\ref{fig:LIFcontour} may present singular points, which can be uncovered, e.g., by examining the functions $P_L$ or $S_L$ more carefully.
As we vary the input $\mu$ for fixed $D$, the LIF model undergoes a transition from the mean- to the noise-driven regime.
In the mean-driven regime (Fig.~\ref{fig:LIFcontour}, panel a), the eigenvalues are complex-conjugate pairs.
For increasing CV, thus allowing for larger fluctuations in the system, the eigenvalues move closer to the real axis (Fig.~\ref{fig:LIFcontour}, panel b).
Eventually, in the noise-driven regime (Fig.~\ref{fig:LIFcontour}, panel c), the eigenvalues become purely real-valued.
In Fig.~\ref{fig:6}, we plot the real and imaginary values of the first eigenvalue $\lambda_1$ for different CVs by varying $\mu$ while keeping $D$ fixed at $D=1/16$, corresponding to the standard deviation $\sigma_V = \sqrt{D} = 1/4$ of the membrane potential $v$.
The transition from mean- to noise-driven dynamics becomes apparent at $C_V \approx 0.85$, where the imaginary part vanishes and the eigenvalue is real-valued.
This transition between mean- and noise-driven dynamics occurs for $\mu$ arguably smaller than $\vth$, whereas the transition from the subthreshold to the suprathreshold regime is at $\mu = \vth$.
The suprathreshold regime, $\mu > \vth$, is characteristic for tonic, i.e., regular spiking behavior.
In the subthreshold regime, $\mu < \vth$, by contrast, noisy fluctuations of the membrane potential $v$ are necessary for a neuron to fire and, thus, the spiking becomes noise-induced \cite{VilLin09}.
For this reason, the smaller $\mu$, the more irregular the spiking and the higher the CV.
However, close to the bifurcation point $\mu \lesssim \vth$ between tonic and noise-induced spiking, the spiking continues to appear regular.
That is why one can expect that the noise-driven dynamics occur only for smaller values $\mu \ll \vth$.
In line, the sub- to suprathreshold transition occurs for $\mu=\vth$ with $\cv = 0.53$. 
By contrast, the transition between the mean- and noise-driven dynamics occurs for smaller $\mu$ and larger CV: $\mu = 0.51 \ll 1 = \vth$ and $\cv \approx 0.85$.

A similar behavior has been found for the PIF neuron with reflecting boundary in \cite{MatGiu02}.
While the eigenvalues are complex-conjugate pairs in the mean-driven regime, they become real in the noise-driven regime. 
This result suggests a general feature of the firing rate dynamics of integrate-and-fire neurons.
For very noisy input and large CV, the relaxation dynamics towards the stationary population activity follows an exponential decay.
By contrast, the mean-driven regime stands out for (damped) oscillatory dynamics as reflected by the complex-valued nature of the eigenvalues.

\subsection{Eigenvalue spectrum for Gaussian distributed ISIs}
\label{subsec:Gaussian}

The Laplace transform of the ISI density can be expressed in terms of its cumulants $\kappa_n(h)$ \cite{Str67I},
\begin{equation}
\label{eq:PLcum}
  P_L(\lambda)=\exp\!\lrrund{\sum_{n=1}^\infty\frac{(-\lambda)^n}{n!}\kappa_n}.
\end{equation}
For tonically spiking neurons, the spike train is regular with a CV much smaller than unity and a Gaussian approximation of the ISI density may be justified. In this case, all ISI cumulants $\kappa_j$ of third and higher order can be neglected, $\kappa_j = 0$ for $j\geq 3$, which simplifies the Laplace transform to
\begin{equation}
\label{eq:PLcum-gauss}
 P_L(\lambda)\approx\exp\!\Big(-\kappa_1\lambda+\frac{\kappa_2}{2}\lambda^2\Big).
\end{equation}
The characteristic equation, Eq.~\eqref{eq:eigenvalue-formula}, becomes
\begin{equation}
\label{eq:kumulamt2}
\frac{\kappa_2}{2}\lambda_n^2-\kappa_1\lambda _n=2\pi i n \ . 
\end{equation}
Solving Eq.~\eqref{eq:kumulamt2} under the constraint that the roots have negative real part, we obtain
%
\begin{align}
\lambda_1&=\frac{\kappa_1}{\kappa_2}\left( 1-\sqrt{1+4\pi i  \frac{\kappa_2}{\kappa_1^2}}\right)\nonumber \\
&=rC_V^{-2}\left( 1-\sqrt{1+4\pi i C_V^2}\right), \label{eq:l1aprox}
\end{align}
where we expressed the rate $r$ and CV in terms of the first two cumulants: $r=\kappa_1^{-1}$ and $\cv=\sqrt{\kappa_2}/\kappa_1$.
In \cite{SchOst13}, Schaffer and co-workers determined the first eigenvalue $\lambda_1$ for different renewal processes empirically by fitting the dependency of the real part and the imaginary part on the rate $r$ and on the coefficient of variation squared $C_V^2$. For small CV, they found the empirical relationship
\begin{equation}
\label{eq:schaffer}
\lambda_1 \approx -r \lreckig{\lrrund{\frac{C_V}{0.22}}^2+2\pi i}. 
\end{equation}
Expanding the square root in Eq.~\eqref{eq:l1aprox} for small CV, we recover a similar expression 
\begin{equation}
\label{eq:l1aprox2}
\lambda_1= -r\left(2\pi^2 C_V^2+2\pi i\right) + \mathcal{O}(C_V^4).
\end{equation}
The numerical evaluation of the factor $1/(\sqrt{2}\pi)=0.225079...$ theoretically explains the empirical result of \cite{SchOst13}.
Note that Eq.~\eqref{eq:l1aprox2} is equivalent to the first eigenvalue of the PIF neuron model, cf.~Eq.~\eqref{eq:PIF_eigenvalues_rateCV}.
The Gaussian approximation, Eq.~\eqref{eq:l1aprox},  presents a valid fit of the first eigenvalue for different renewal processes for low $C_V\leq 0.3$ (Fig.~\ref{fig:6}, black line). 
For larger CV, the real parts as well as the imaginary parts of the first eigenvalue for the different models deviate from the Gaussian approximation.
The absolute value of the real part, which represents the dominant relaxation rate towards the stationary solution in the firing rate model Eq.~\eqref{eq:first-mode-approx}, continually grows for the PAR, PIF and LIF models, whereas the Gaussian approximation saturates. In contrast, the Gamma model exhibits a non-monotonous behavior: While the real part of the first eigenvalue decreases for intermediate CV, it moves closer to the imaginary axis for large CV. 
The imaginary part of the first eigenvalue prescribes the frequency of oscillatory transients.
In the PIF model, the first eigenvalue has constant imaginary part $\mathrm{Im}(\lambda_1)=2\pi r$ when varying the CV.
This value coincides with the other three models and with  the Gaussian approximation for small $C_V\leq 0.2$.
For larger CV, the frequency increases for the PAR model but decreases for the Gamma model and for the Gaussian approximation.

\section{Comparison of the approximate low-dimensional firing rate model with the full population density model}
\label{sec:low-dim_pop_dyn}

\begin{figure*}[t]
\begin{center}
\includegraphics[scale=1]{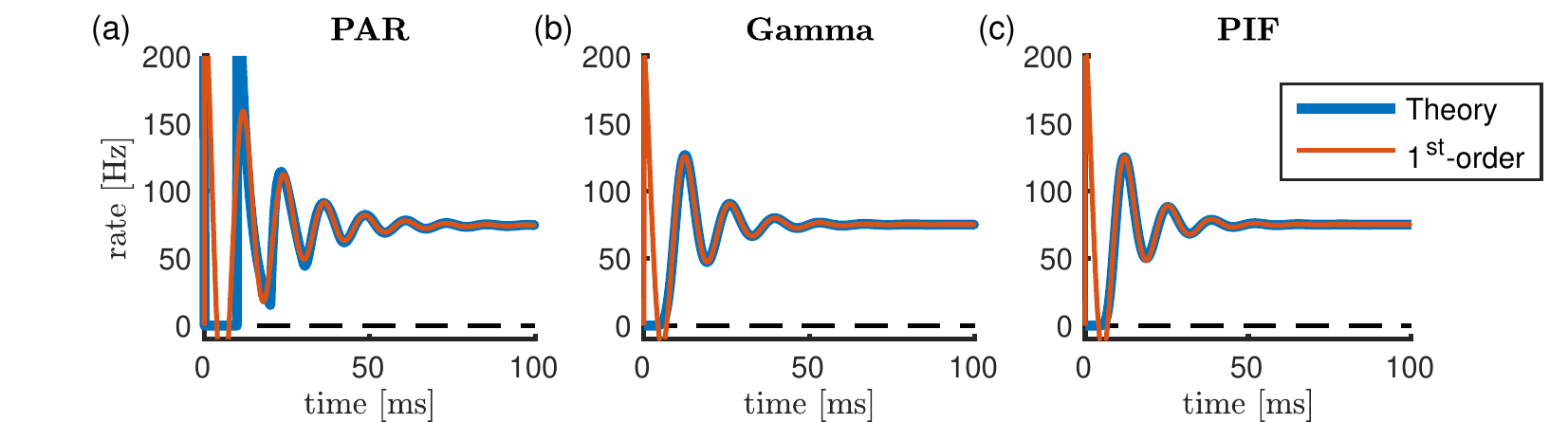}
\caption{Peristimulus time histogram (PSTH) of uncoupled populations of PAR, Gamma, and PIF neurons. At time $t=0^-$, all neurons fire synchronously and are subsequently reset, corresponding to $a_1(0)=1$. Then, a step-input current $h$ is injected such as to asymptotically drive all neurons to fire with mean rate $r=75$~Hz and with a CV of $\cv=1/\sqrt{15}$.
All three neuron models capture the effect of refractoriness after the initial population spike.
Note that the PAR and Gamma models are fully determined by $r$ and $\cv$. For the PIF model we additionally used $\vth=10$ mV and $\mu(h)=h$.
}
\label{fig:3}
\end{center}
\end{figure*}

To evaluate the performance of the low-dimensional firing rate model, Eq.~\eqref{eq:first-mode-approx}, we compare the time-dependent behavior of this model with the corresponding prediction of the exact refractory density dynamics. In particular, we consider three types of dynamical behavior: (i) transient relaxation dynamics to equilibrium for constant input $h$ (as occurring for a step input); (ii) dynamical response to weak time-dependent external stimuli; and (iii) emergent oscillatory dynamics in a population of recurrently connected neurons corresponding to partial synchronization. 

\subsection{Transient dynamics of populations of non-interacting neurons}
We first consider the transient population dynamics of uncoupled neurons in the case when all neurons are initially synchronized by imposing that all neurons have just fired a spike at $t=0$ (Fig.~\ref{fig:3}). For $t>0$, we assume a constant input current $h$ and parameters such that different models have identical stationary firing rate 
and coefficient of variation (CV) in its long-time behavior for $t\rightarrow\infty$. 
The first-order approximation, Eq.~\eqref{eq:first-mode-approx}, shows perfect agreement with the theoretical prediction obtained from the refractory density equation, Eq.~\eqref{eq:popact-integr}.
Because of the complex-valued eigenvalues, the ringing behavior into the stationary solution is recovered, which is generically not possible in classic rate models, see, e.g., \cite{SchOst13,DevRox17}.
%
%

For non-interacting neurons ($J=0$) and constant input $h$ for $t>0$, the firing-rate model Eq.~\eqref{eq:first-mode-approx} simplifies to
\begin{align}
  \label{eq:fr-simple}
  A(t)&=F_0(h) + 2 \mathrm{Re}\big[ a_1(t) F_1(h) \big],\\
  \od{a_1}{t}&=\lambda_1(h) a_1,\qquad a_1(0)=1.  
\end{align}
To evaluate this dynamics, we only have to identify the first eigenvalue $\lambda_1$ from the eigenvalue formula Eq.~\eqref{eq:eigenvalue-formula}, $P_L(\lambda_1)=1$, the transfer function $F_0(h)$ and the factor $F_1(h)$, which can be readily computed for the examples presented in Sec.~\ref{sec:spec_decomp_specific}.
The first eigenvalue $\lambda_1$ crucially determines the response dynamics to a constant input current.
The real part represents the main relaxation rate towards the stationary solution and therefore describes the envelope of the oscillatory transient dynamics in Fig.~\ref{fig:3}.
The main frequency of the decaying oscillations is given by the imaginary part of the first eigenvalue.
The agreement between the exact population dynamics and the first-order approximation supports the view that relaxation rate and frequency of the oscillatory response are well captured by the first eigenvalue $\lambda_1$ and higher modes have only a negligible effect on the accuracy.

\subsection{Population dynamics in response to weak modulation of input current}
\label{sec:popul-dynam-resp}

\begin{figure*}[t]
\begin{center}
\includegraphics[scale=1]{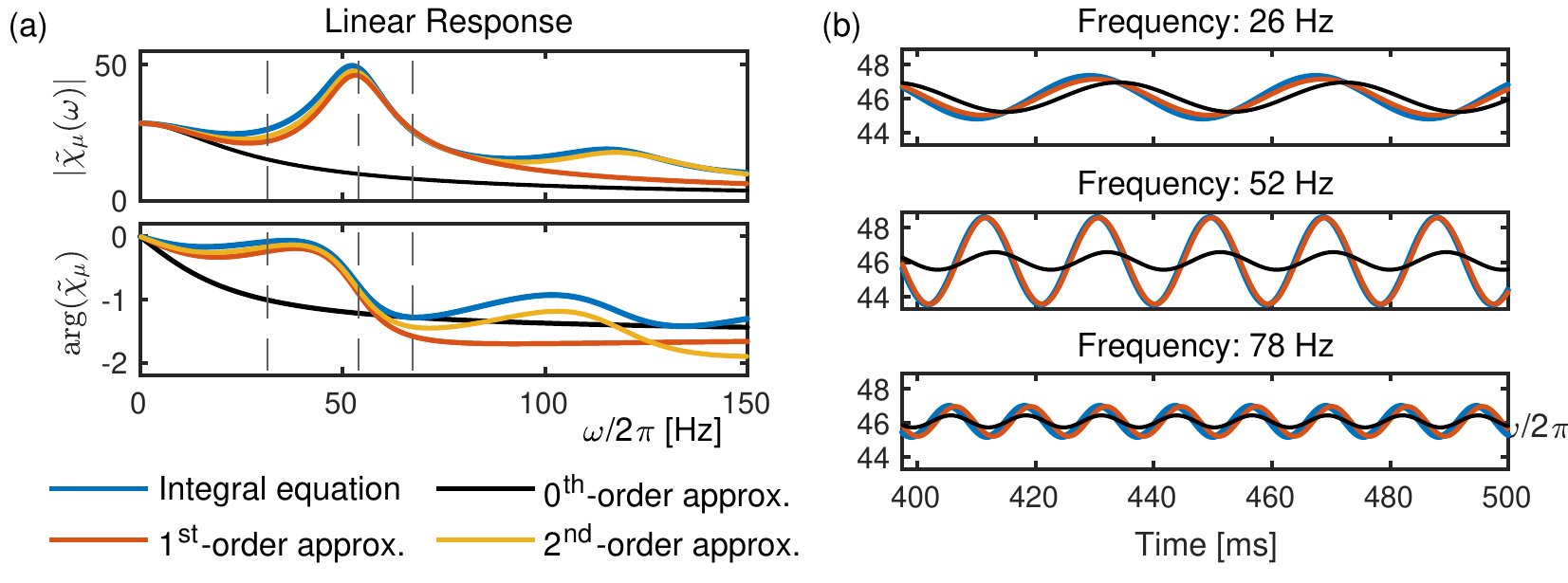}
\caption{(a) Linear response of an uncoupled population of Poisson neurons with absolute refractoriness (PAR model).
Top: the absolute value, bottom: the phase of the linear response to an additional sinusoidal input $\mu$ on top of the external input current $I(t) = I_0 + \mu(t)$.
The exact linear response (blue) is obtained with the integral equation \eqref{eq:Pois_integraleq}.
While the zeroth-order approximation (black) represents only a low-pass filter and does not capture any resonance, the first-order approximation Eq.~\eqref{eq:first-mode-approx} (red) accurately captures the first resonance peak and the second-order approximation (yellow) even the first two resonances. 
(b) Trajectories subject to an oscillatory drive at three exemplary frequencies follow the theoretical predictions for the integral equation, zeroth- and first-order approximations, with amplitude and phase deviations as indicated by the dashed vertical lines on the left.
Parameters: $\Delta=15$~ms, $\tau_h=8$~ms, $J=0$~mV/kHz, $\nu_0=100$~Hz, $\delta=0.5$~mV,  $\theta=1$~mV and  $I_0=1.2$~mV.
}
\label{fig:4}
\end{center}
\end{figure*}

Next, we consider a weak time-dependent stimulus $I(t)=I_0+\mu(t)$, where $I_0$ is a constant base current and $\mu(t)\ll I_0$ is a weak modulation. For simplicity, we still assume an uncoupled population ($J=0$) here; hence,  $h(t)=h_\text{ext}(t)=(\kappa*I)(t)$. For concreteness, we consider an exponential filter function  $\kappa(t)=\tau_h^{-1}e^{-t/\tau_h}\Theta(t)$ corresponding to the first-order kinetics
\begin{equation}
  \label{eq:h-first-ord}
  \tau_h\od{h}{t}=-h+I_0+\mu(t),
\end{equation}
as in classical firing rate models. For arbitrary (but weak) stimuli $\mu(t)$, the time-dependent response of the population activity can be compactly represented by the linear response function $\chi_\mu(t)=(\kappa*\chi_h)(t)$:
\begin{equation}
  \label{eq:lin-response}
A(t)=A_0+A_1(t)=F_0(I_0)+(\chi_\mu*\mu)(t).  
\end{equation}
Because of linearity of Eq.~\eqref{eq:lin-response}, it is sufficient to regard weak periodic stimuli of the form $\mu(t)=\varepsilon \cos(\omega t)$. In this case, the response of the population activity is cosinusoidal with the same angular frequency $\omega$,
\[A_1(t)=\varepsilon|\tilde\chi_\mu(\omega)|\cos(\omega t+\phi(\omega)),\qquad \phi(\omega)=\arg(\tilde{\chi}_\mu(\omega)),\]
where the frequency-dependent amplitude and phase shift are given, respectively, by the modulus and the argument of linear response function $\tilde{\chi}_\mu(\omega)=\tilde{\kappa}(\omega)\tilde{\chi}_h(\omega)$ in frequency space. The linear response function $\tilde{\chi}_h(\omega)$ is given by Eq.~\eqref{eq:lin-respon} for the low-dimensional firing rate model (first-order approximation, Sec.~\ref{sec:low-dimens-firing}) and $\tilde{\kappa}(\omega)=(1+i\omega\tau_h)^{-1}$. 

To evaluate the performance of the first-order approximation for arbitrary stimulus frequencies $\omega$, we compare the linear response function to the exact linear response function  as well as to the linear response function of the classical firing rate model with the same steady-state properties (zeroth-order approximation, Eq.~\eqref{eq:zeroth-order}). In principle, the exact linear response can be obtained from the linearization of the refractory density equation. As an example, we here use the Poisson model with absolute refractoriness (PAR model, Sec.~\ref{sec:poisson-model-with}), because it allows for a closed-form analytical expression of $\tilde\chi_h(\omega)$ (see Appendix, Sec.~\ref{sec:linearizeddyn-integr-par}), and hence also of $\tilde\chi_\mu(\omega)$. 
The linear response function of the zeroth-order approximation Eq.~\eqref{eq:zeroth-order} merely reflects the filter function $\kappa(t)$ because $\tilde\chi_h(\omega)=F_0'(I_0)$, where the steady state transfer function $F_0$ is given by Eq.~\eqref{eq:F0-par}. Because of the frequency-independence of $\tilde{\chi}_h(\omega)$ and the low-pass behavior of $\tilde{\kappa}(\omega)$, the zeroth-order model does not exhibit resonances as exemplarily shown in Fig.~\ref{fig:4}.
The linear response of the first-order approximation, Eq.~\eqref{eq:lin-respon}, by contrast, does capture the main resonance peak with good agreement in both its amplitude as well as its phase when compared to the true linear response function. This agreement can also be seen in the exemplary trajectories with different driving frequencies (Fig.~\ref{fig:4}b). By construction, resonances at higher harmonics of the fundamental frequency cannot be captured by the zeroth- nor the first-order approximation.
By adding higher modes in the low-dimensional firing rate model, however, it is possible to also capture these higher resonance peaks. For example, the second-order model accurately describes the frequency response at the (first) dominant frequency as well as at the second harmonics (Fig.~\ref{fig:4}a).
%
%

\subsection{Population dynamics of interacting neurons}
\label{sec:popul-dynam-inter}

%
\begin{figure*}[t]
\begin{center}
\includegraphics[scale=1]{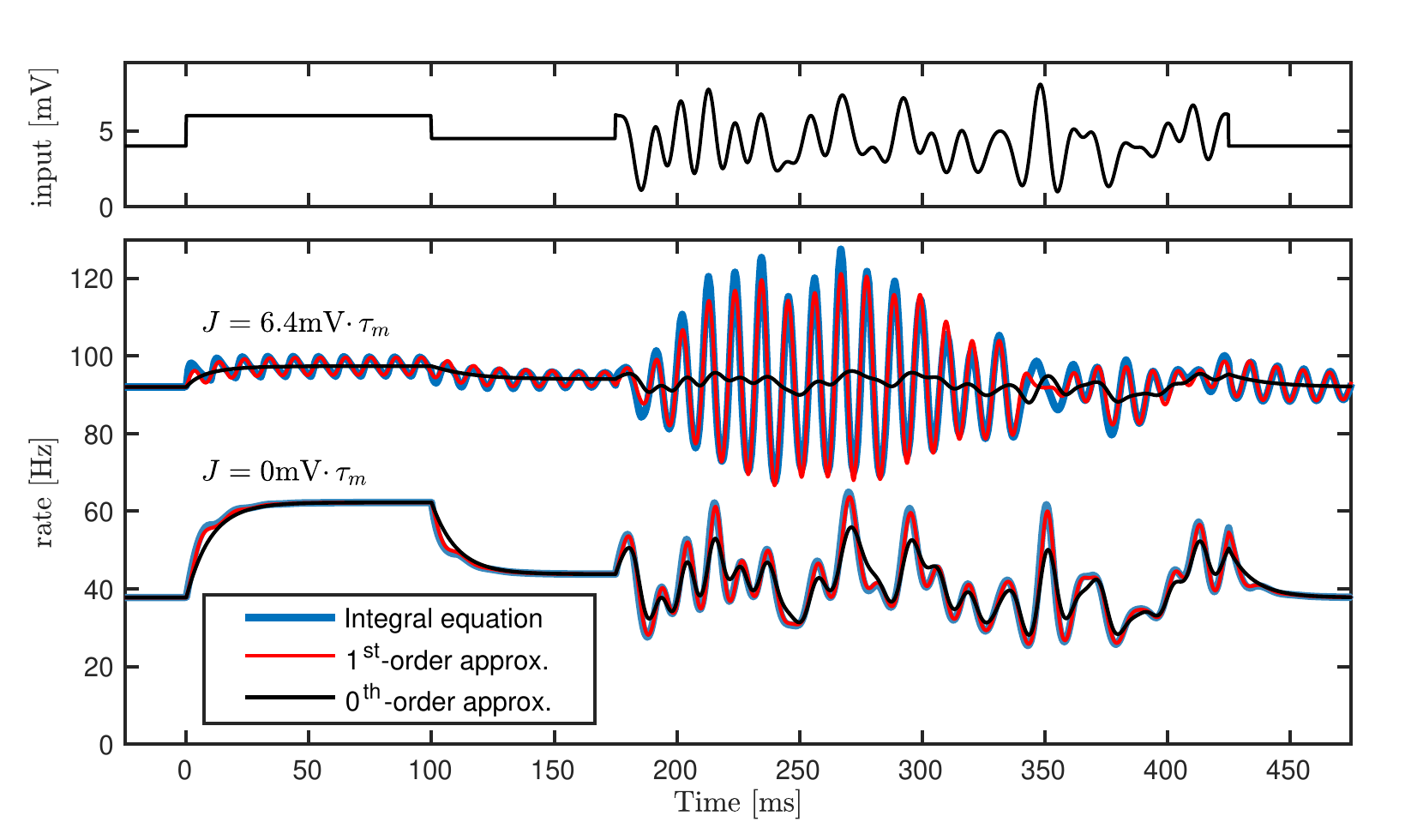}
\caption{Recurrent population of excitatory PAR neurons subject to an external input current $I(t)$ (top).
The collective oscillations of the population activity (bottom) are due to spike synchronization.
While the first mode approximation (red) follows the theoretical predictions (blue), the population of effective Poisson neurons (black) cannot capture spike synchrony effects.
Parameters are $\Delta=5$ ms, $\kappa(t) = \tau_h^{-1} \exp(-t/\tau_h)$ and $\alpha(t) = \tau_s^{-1} \exp(-t/\tau_s)$ with membrane and synaptic time constants $\tau_h=20$ ms and $\tau_s=10$ ms, respectively, such that $\epsilon(t) = (\kappa \ast \alpha)(t)$, $\nu_0 = 100$ Hz, $\delta = 2$ mV, $\theta=5$ mV, and $J$ as indicated.
}
\label{fig:5}
\end{center}
\end{figure*}

Allowing for recurrent coupling among neurons, may result in collective oscillations due to (partial) spike synchronization.
Linear response theory can also be used in this case \cite{BruWan03,GerKis14}. 
An oscillatory instability of the population activity occurs for a critical coupling strength $J_c$ with emerging frequency $\omega_c/(2\pi)$ if
an amplitude condition ($|J_c \tilde\epsilon(\omega_c)\tilde{\chi}_h(\omega_c)|=1$) and a phase condition ($\text{arg}\big(J_c \tilde\epsilon(\omega_c)\tilde\chi_h(\omega_c)\big) = 0\mod 2\pi$) are met; $\tilde{\chi}_h$ denotes the Fourier transform of the linear response function $\chi_h$ with respect to changes in the input current $h(t)$ (see Appendix, Sec.~\ref{sec:stab-analys-asynchr}).
While the first condition can always be met for sufficiently strong coupling strength $J\geq J_c$, the latter requires a balance of the various time scales in the system, such as of the absolute refractory period $\Delta$ and the synaptic dynamics with filter function $\epsilon(t)$.
Note that we only need the linear response function to find the bifurcation point, so that this approach can directly be applied to the first-order approximation Eq.~\eqref{eq:first-mode-approx}, e.g., when the integral equation is analytically intractable.
In an exemplary recurrent network of excitatory PAR neurons, we are thus able to find the critical coupling strength $J_c$, which will, in general, depend on the applied current $I(t)$.
Beyond the critical point, the stationary solution $A_0$ of the population activity becomes unstable and gives rise to collective oscillations.
In Fig.~\ref{fig:5}, we illustrate the onset of synchronization by varying the input $I(t)$.
Without recurrent coupling, $J=0$, the zeroth-order approximation Eq.~\eqref{eq:zeroth-order} (Fig.~\ref{fig:5}, black curve) is still able to approximate the full dynamics (blue) sufficiently well although, for noisy, quickly varying input, it can only follow the trends but not the full transient behavior.
Incorporating the first mode (red), though, leads to a perfect agreement with the full dynamics as already seen for the peristimulus time histogram (PSTH) in Fig.~\ref{fig:4}.
With sufficient recurrent coupling, $J>0$, an increase in input $I(t)$ makes it possible that the afore-mentioned amplitude and phase conditions are fulfilled and collective oscillations occur at time $t=0$ in Fig.~\ref{fig:5}.
The classical rate model as given by the zeroth-order approximation Eq.~\eqref{eq:zeroth-order} cannot account for these oscillations nor for the spike synchronization effects due to the complex input (for $t>175$ ms).
By contrast, the first-order approximation Eq.~\eqref{eq:first-mode-approx}  captures both of these complex dynamics. It allows for oscillatory population activity and also follows the correct transient response due to fast fluctuating inputs.




\section{Discussion}
\label{sec:discussion}

In this paper, we used the spectral method to derive a system of ordinary differential equations that represents a principled firing rate (FR) model for macroscopic populations of spiking neurons. The method works for spiking neurons modeled as arbitrary, time-dependent renewal processes including neuron models with pronounced refractoriness. By discarding higher-order eigenmodes and retaining the most dominant ones, this approach permits a systematic way to obtain low-dimensional FR models. Interestingly, we uncovered a simple relation that links the dominant time scales of the macroscopic population dynamics with the interspike interval density of single neurons. We demonstrated the method for simple renewal models, for which the eigenvalues can be calculated analytically. We observed that the low-dimensional FR model based on the first mode (first-order approximation) already captures fast transients, linear response properties (apart for higher harmonics) and spike synchronization dynamics sufficiently well. As a by-product, our theory also provides an explanation of a previously discovered empirical formula \cite{SchOst13} for the dominant eigenvalue in terms of the rate and CV, which is valid for low CV $\leq0.3$ (see \emph{Section}~\ref{subsec:Gaussian}).


Homogeneous populations of renewal neurons are widely used in theoretical neuroscience \cite{AbbVre93,Bru00,PleGer00,LinDoi05,PotDie14} because they lead to compact population density equations \cite{NykTra00,OmuKni00,ChiGra07} for a single state variable (e.g. membrane potential or neuronal age). However, population density equations are partial differential equations or integral equations and as such they represent an infinite-dimensional dynamics. In contrast to heuristic FR models which are low-dimensional, population density approaches are thus much less efficient for numerical integration and, although possible, they are much harder to tackle analytically.
Therefore, our reduction of the refractory density equation to low-dimensional FR models may be useful for both efficient numerical simulations and analytical studies.

Regarding numerical efficiency, a low-dimensional reduction is particularly critical for systems containing a large number of populations such as spatially extended field models discretized on a grid, recurrent neural networks of populations that can be trained to perform a task \cite{SusAbb09}, attractor networks for associative memory \cite{GerHem92b}, networks with many assemblies (clusters) \cite{LitDoi12} or cortical circuits containing many cell types per cortical layer \cite{BilCai20}. In these cases, if neurons are modeled as renewal spiking neurons, each population would possess its own refractory density equation making the integration of the multi-population system computationally expensive. In contrast, approximating the dynamics of each population by the low-dimensional FR model derived in this paper would dramatically speed up the numerical integration of the multi-population dynamics while maintaining the effect of refractoriness.

Regarding analytical studies, the low-dimensional firing rate model may enable progress for multi-population systems of spiking neurons, for which traditional population density approaches would be infeasible. For example, there has been recent advances to determine the intrinsically generated chaotic variability in large recurrent networks of multi-dimensional firing rate units \cite{MusGer19}. With the multi-dimensional rate models derived for spiking neurons in the present paper, it may be possible to obtain a similar self-consistent mean-field theory of intrinsically generated fluctuations for recurrent networks of spiking neurons.

The reduction of the infinite-dimensional refractory density equation to low-dimensional FR models is also a crucial step towards a mathematical framework for cortical circuit models. In an interesting line of research, detailed biophysical circuit models \cite{MarMul15,ArkGou18} and neuronal recordings have been reduced or fitted to generalized integrate-and-fire (GIF) point neuron models \cite{RosPoz16,PozMen15,TeeIye18,BilCai20}; in turn, cortical circuits of GIF neurons have been reduced to mesoscopic circuit models based on an extension of the refractory density equation to finite-size neuronal populations  \cite{SchDeg17}. If we were able to further simplify the mesoscopic cortical circuit model to a low-dimensional FR model, this would provide  opportunities to study emergent dynamical behavior of cortical circuits that can be traced back to the underlying biophysics at the neuronal level. For a recent opinion on the use of the refractory density framework for modeling cortical dynamics, see \cite{SchChi19}.

\subsection{Extensions}
\label{sec:extensions}

Eventually, our goal is to develop a new class of firing-rate models for mesoscopic cortical activity that can be constrained by physiological parameters and captures essential features of cortical dynamics and variability. In this paper, we have shown how to incorporate basic  features of spiking dynamics such as refractoriness and spike-synchronization into low-dimensional FR models. Several other important features are however still missing: \emph{First}, renewal models do not exhibit \emph{spike-frequency adaptation} \cite{SchLin13} -- a basic property of many cell types \cite{PozNau13}. Slow adaptation mechanisms can be treated by a mean-adaptation approximation \cite{BenHer03,LaCRau04,Sch13}, in which the adaptation variable is effectively driven by the population activity. The present theory can be extended to this case in a straightforward manner because the resulting mean adaptation can be treated as one element of a parameter vector $\vec{h}(t)$, as shown in \cite{GigMat07}. Adaptation has also been treated by the quasi-renewal (QR) approximation \cite{NauGer12}, however, it is less evident how to extend the eigenfunction method to the QR system.

\emph{Second}, we assumed an infinitely large (``macroscopic'') population of neurons, and hence neglected \emph{finite-size fluctuations}. However, a realistic model of cortical dynamics should account for finite-size effects at the mesoscopic scale. For example, in mouse barrel cortex, neuron numbers have been estimated to be only on the order of 100 to 1000 per neuron type and cortical layer of a barrel column \cite{LefTom09}, which implies substantial finite-size fluctuations \cite{SchChi19}. Previous theoretical studies \cite{MatGiu02,MatGiu04} have incorporated finite-size noise using a similar eigenfunction approach as studied here. However, those studies were based on a heuristic extension of the Fokker-Planck equation to finite population size that breaks the conservation of neurons. In particular, the heuristic extension does not correctly capture the effect of refractoriness on finite-size fluctuations, especially at low frequencies. Therefore, a low-dimensional firing rate model for finite-size populations that correctly predicts the low-frequency behavior of finite-size fluctuations is still lacking. In contrast, a principled finite-size extension has recently been achieved for  the refractory density equation \cite{SchDeg17}. This extension respects the conservation of neurons and accurately describes finite-size fluctuations at all frequencies. The mesoscopic dynamics is of the form of a stochastic integral equation, and as such, it is still infinite-dimensional. However, combining this theory with the low-dimensional reduction of the refractory density equation presented here for the macroscopic case $N\rightarrow\infty$, will enable a systematic derivation of low-dimensional stochastic dynamics for the mesoscopic case $N<\infty$.    

\emph{Third}, we assumed static synapses with constant weights $J$, whereas cortical synapses are often dynamic exhibiting \emph{synaptic short-term plasticity} (STP). In a recent paper \cite{SchGer18arxiv}, we have developed an extension of the mesoscopic population dynamics \cite{SchDeg17} that includes STP. It has been shown that this extension works for heuristic FR models (corresponding to the zeroth-order approximation in this paper) as well as for accurate stochastic integral equations for populations of spiking neurons. Thus, we expect that the STP extension can be applied in a straightforward manner to the first- or higher-order FR models that developed here as well. And \emph{fourth}, we assumed that populations are approximately homogeneous, while cortical circuits exhibit \emph{heterogeneity}. How to deal with heterogeneity in a  mean-field approach is a crucial question for future research. For example, heterogeneity in neuronal parameters may counteract spike synchronization. The relative contribution of neuronal dynamics and heterogeneity to an effective FR model remains, however, unclear.

\subsection{Open questions}
\label{sec:open}

Finally, we want to mention some open technical questions that are beyond the scope of the current paper. First, the eigenvalue formula Eq.~\eqref{eq:eigenvalue-formula} to compute the first eigenvalue, $P_L(\lambda_1)=1$, may be difficult to interpret if the Laplace transform $P_L(\lambda)$ is not known analytically for complex arguments $\lambda$. In this case, we are looking for some $\lambda\neq 0$ that leaves the integral $\int \mathrm{e}^{-\lambda \tau }P(\tau)d\tau = 1 = \int P(\tau) d\tau$ unchanged, where we used that the ISI distribution $P(\tau)$ is normalized. How to find such $\lambda$ even numerically remains an open problem.

Second, the computation of the coupling coefficients $c_{nm}(h)$ and $\hat{c}_{nm}(h)$ in Eq.~\eqref{eq:an-dyn} is challenging except for special cases such as the Poisson process with absolute refractoriness. \citet{SchOst13} have omitted the coupling terms stating that 
they ``are typically small enough to be ignored.''
A systematic investigation how the coupling coefficients affect the low-dimensional dynamics is still pending.
Preliminary results hint at a hierarchy similar to that of the $a_{n}$, see also \cite{Mat16_arxiv}, which underlines the assumption that higher modes have a negligible impact on lower modes.
Keeping the coupling terms in the first-order approximation Eq.~\eqref{eq:first-mode-approx}, however, proved crucial to correctly replicate the population activity in Fig.~\ref{fig:5} of the present study. In particular, taking the coupling terms into account allowed us to capture the spike synchronization effects and collective oscillations in the recurrent network of spiking neurons. Thus, an important task for further studies is to obtain simple approximations for the coupling coefficients in cases where these coefficients cannot be calculated exactly.

Third, the gamma model with non-integer shape parameter $\alpha$ may be of interest for practical applications, however, this case is not yet well understood. For example, \citet{Ost11} investigated the ISI densities of various integrate-and-fire models as well as conductance-based models and concluded that ``the first two moments essentially determine the full ISI distribution independently of the model considered.''
\citet{ShiSak99} further showed that the CV and the skewness (SK) of the ISI distribution of leaky integrate-and-fire dynamics are constrained to a narrow region in the CV-SK plane close to SK $=2$ CV.
This line corresponds directly to the Gamma distribution \cite{SchMik13}, which has frequently been used to fit ISI distributions experimentally measured from cortical neurons, see, e.g., \cite{BarQui01,MiuTsu07,MaiAss09}. It is tempting to use the fitted parameters of the Gamma distribution in the low-dimensional FR model studied in Sec.~\ref{sec:gamma-model} because there is an explicit analytic formula for the eigenvalues $\lambda_n$ of the Gamma model. Such a continuous extension of the Gamma model could be directly determined from experimentally measured rates and CV's of cortical cells. 
However, a fitted shape parameter $\alpha$ would generally be non-integer. An extension of the eigenvalue formula Eq.~\eqref{eq:gammaeig} to non-integer shape parameters raises some interesting questions: e.g., what is the range of $n$ if, e.g., $\alpha<2$? Is $n$ still bounded by $\alpha-1$? Is the eigenvalue spectrum $\{\lambda_n\}$ discrete or continuous if $\alpha$ is irrational? Does the model allow to model bursty neurons corresponding to a shape parameter $\alpha<1$? These questions remain to be answered in future studies.   
However, Fig.~\ref{fig:6} also shows that the dependence of the dominant eigenvalue on rate and CV is not as model-independent as suggested by \citet{SchOst13}. On the contrary, our study shows that the dominant eigenvalue of different neuron models may strongly deviate from the eigenvalue of the gamma model with the same rate and CV. This result indicates that the information of rate and CV is not sufficient to approximately determine the eigenvalue spectrum (e.g. by using the corresponding formulas for the Gamma model). An interesting question will thus be whether the eigenvalue can be constrained by further pieces of information such as the skewness of the ISIs.

In conclusion, given the broad class of neurons that can be described as time-dependent renewal processes or quasi-renewal process, our novel FR model opens the way for modeling cortical population dynamics based on spiking neuron models with realistic refractoriness.

\section{Acknowledgments}
TS and NG would like to thank Wulfram Gerstner for his wide support during part of the research. We also thank Sven Goedeke for fruitful discussions, especially about the derivation in Appendix, Sec.~\ref{sec:alt-deriv}. 


\newpage
\appendix


\section{Derivation of Eqs.~\eqref{eq:an-dyn} and \eqref{eq:popact-an}}
\label{sec:append-a:-deriv}
Here, we review the general concept of the spectral decomposition method for the population dynamics of neurons characterized by a one-dimensional state variable $x$ and a (possibly multi-dimensional) control parameter $\vec{h}=[h_1,\dotsc,h_M]^T$. We assume that the population density $p(x,t)$ satisfies an evolution equation of the form $\partial_tp=\op{L}_xp$, where $\op{L}_x$ is a linear evolution operator. In the context of the refractory density equation (\ref{eq:refr-dens}~\!\&\!~\ref{eq:popact-integr}), the state variable is the age $x=\tau$. The derivation has been developed for the Fokker-Planck equation for the membrane potential density $p(v,t)$ \cite{KniMan96,MatGiu02}, where the state variable is the membrane potential $x=v$. 
First, we consider the eigenfunctions $\{\phi_n(x,\vec{h})\}$ and $\{\psi_n(x,\vec{h})\}$ of the operator $\op{L}_x(h)$ and its adjoint operator $\op{L}_x^+(h)$, respectively,
\begin{align}
  \label{eq:eigenfuncs}
  \op{L}_x(h)\phi_n(x,\vec{h})&=\lambda_n\phi_n(x,\vec{h})\\
  \op{L}_x^+(h)\psi_n(x,\vec{h})&=\lambda_n^+\psi_n(x,\vec{h})
\end{align}
with respect to the scalar product $\langle f|g\rangle:=\int f(x)g(x)\,dx$ of two functions $f(x)$ and $g(x)$.
The adjoint operator $\op{L}_x(h)$ has then the same eigenvalues as $\op{L}_x(h)$, i.e. $\lambda_n^+=\lambda_n$. Furthermore, after appropriate normalization, the eigenfunctions form a  bi-orthonormal basis such that
\begin{equation}
  \label{eq:bi-orth}
  \langle\psi_n|\phi_m\rangle=\delta_{n,m},
\end{equation}
where $\delta_{n,m}$ denotes the Kronecker delta function.
As in the spectral decomposition for Fokker-Planck systems, we expand the population density into eigenfunctions of the evolution operator $\op{L}_x$:
\begin{equation}
  \label{eq:app_spec-decomp}
  p(x,t)=\sum_na_n(t)\phi_n(x,\vec{h}).
\end{equation}
The complex variables $a_n(t)$ are the projections of $p(x,t)$ on the $n$-th eigenmode: 
\begin{equation}\label{eq:appA_an_def}
a_n(t)=\langle \psi_n | p \rangle.
\end{equation}
As the eigenvalues of the real-valued non-Hermitian operator $L_x$ come in complex-conjugates pairs, $\lambda_n, \lambda_n^*$, it follows from Eq.~\eqref{eq:eigenfuncs} that 
\begin{equation} \label{eq:phin_symmetry}
 \phi_{-n} = \phi_n^* \quad \text{ and } \quad  \psi_{-n} = \psi_n^* \ 
\end{equation}
are the eigenfunctions of $\op{L}_x$ and $\op{L}_x^+$ associated with the eigenvalue $\lambda_{-n}:=\lambda_n^*$.
Consequently, we also have that
\begin{equation} \label{eq:an_symmetry}
a_{-n}(t)=\langle \psi_{-n} | p \rangle=\langle \psi_{n}^* | p \rangle = \langle \psi_{n} | p \rangle^* = a_n^*(t)
\end{equation}
because $p(\tau,t)$ is real-valued.

The dynamics of the $n$th mode can be found as
\begin{align}
    \od{a_n}{t}&=\langle\psi_n|\partial_tp\rangle+\langle\partial_t\psi_n|p\rangle\\
             &=\langle\psi_n|\op{L}_x p\rangle+\od{\vec{h}}{t}\cdot\left\langle\partial_{\vec{h}}\psi_n\middle|p\right\rangle \nonumber \\
&=\langle\op{L}_x^+\psi_n| p\rangle+\od{{\vec{h}}}{t}\cdot\sum_{m\in\mathbb{Z}}a_m\langle\partial_{\vec{h}}\psi_n|\phi_m\rangle \nonumber \\
&=\lambda_n\langle\psi_n|p\rangle+\od{{\vec{h}}}{t}\cdot\sum_{m\in\mathbb{Z}}a_m\langle\partial_{\vec{h}}\psi_n|\phi_m\rangle \nonumber \\
&=\lambda_na_n+\od{{\vec{h}}}{t}\cdot\sum_{m\in\mathbb{Z}}a_m\vec{c}_{nm}, \label{eq:deri-andot}
\end{align}
where we introduced the coupling coefficients
\begin{equation}
\label{eq:cnmcoefficents}
\vec{c}_{nm}=\langle\partial_{\vec{h}}\psi_n|\phi_m \rangle.
\end{equation}
Alternatively, we can express the sum in Eq.~\eqref{eq:deri-andot} in terms of modes $a_m$ with only positive indices $m>0$,
\begin{equation}
\label{eq:andyn-simp}
\vec{c}_{n0}+\sum_{m=1}^\infty \big(\vec{c}_{nm}a_m+\vec{c}_{n(-m)}a_m^*\big),
\end{equation}
where we used Eq.~\eqref{eq:an_symmetry}.

The initial conditions $a_n(0)$ of the modes $a_n(t)$ depend on the initial refractory density $p(\tau,0)$ and can be determined through Eq.~\eqref{eq:appA_an_def} as 
$a_n(0)=\langle \psi_n(\tau,\vec{h}) | p(\tau,0) \rangle$.
Starting from the stationary state $p(\tau,0)=\phi_0(\tau,\vec{h})$, 
the initial conditions are $a_n(0)= \langle \psi_n | p(\tau,0) \rangle = \langle \psi_n | \phi_0 \rangle = \delta_{n,0}$, where we used the orthonormality relation Eq.~\eqref{eq:bi-orth}.
Starting from the fully synchronized state, in which all neurons had a spike immediately before at time $t=0^-$, the initial refractory density is $p(\tau,0)=\delta(\tau)$.
Hence, the initial conditions are $a_n(0) = \langle \psi_n(\tau,\vec{h}) | \delta(\tau) \rangle = \psi_n(0,\vec{h}) = 1$  because of the normalization $\psi_n(0,\vec{h}) = 1$ for all $n\in\mathbb{Z}$.

Let us consider the operator $\op{L}_x^\text{fire}$ that extracts the firing rate (resp. population activity) from the current population density: $A(t)=\op{L}_x^\text{fire}p(x,t)$. Using the expansion Eq.~\eqref{eq:app_spec-decomp}, we find in general
\begin{equation}
\label{eq:popact-gen-app_modeapprox}
A(t)=\sum_{n\in\mathbb{Z}}a_n(t)\op{L}_x^\text{fire}\phi_n(x,\vec{h})=F_0(\vec{h})+2\sum_{n=1}^\infty\text{Re}\big(F_n(\vec{h})a_n\big)
\end{equation}
where we abbreviated $F_n(\vec{h}) = \op{L}_x^\text{fire}\phi_n(x,\vec{h})$ and used the symmetries Eqs.~\eqref{eq:phin_symmetry} and \eqref{eq:an_symmetry}.

In our case of the refractory density equation, where the state variable is the age $x=\tau$, the population activity is given by the refractory density at zero age, $A(t)=p(0,t)$, and hence the firing rate operator is defined by the action on a function $f(\tau)$ as
\begin{equation}
  \label{eq:fire-op}
  \op{L}_\tau^\text{fire}f(\tau)=f(0).
\end{equation}
This definition yields Eq.~\eqref{eq:popact-an} in the main text, in particular, $F_n(\vec{h})=\phi_n(0,\vec{h})$. As an aside, we mention that in the Fokker-Planck formalism for IF models driven by white Gaussian noise, Eq.~\eqref{eq:ifdyn_general1}, the firing rate arises from the derivative of the membrane potential density at the threshold, $A(t)=-D\partial_vp(\vth,t)$. Thus, the firing rate operator would read in that case
\begin{equation}
  \label{eq:fpe-fire-op}
  \op{L}_v^\text{fire}f(v)=-D\partial_vf(\vth),
\end{equation}
and thus $F_n(\vec{h})=-D\partial_v\phi_n(\vth,\vec{h})$.


  

\section{Linear response theory}

\subsection{Linearized dynamics of eigenmodes and linear response function}
\label{sec:linearizeddyn-eig-append}

Let us assume that for a constant control parameter $h(t)=h_0$ the population activity is stationary $A(t)=A_0=F_0(h_0)$. We are interested in the dynamics close to the stationary activity if the input is weakly modulated, $h(t)=h_0+\varepsilon h_1(t)$, with $\varepsilon\ll 1$. In the spectral representation of the dynamics, Eq.~\eqref{eq:an-dyn}, the weak modulation $\varepsilon h_1(t)$ induces small amplitudes $a_n(t)=\varepsilon a_{n,1}(t)+\mathcal{O}(\varepsilon^2)$, $n=1,2,\dotsc$, of order $\varepsilon$ (assuming that initial conditions $a_n(0)$ are at most of order $\varepsilon$). Taylor expansion and keeping only zero- and first-order terms yields the linearized dynamics      
\begin{subequations}\label{eq:mode-approx-lin}
\begin{align}
  A(t) &= F_0\big(h_0\big)+\varepsilon A_1(t),\\
  A_1(t)&=F_0'(h_0)h_1(t) +  2\sum_{n=1}^\infty\mathrm{Re} \big[ F_n(h_0)a_{n,1}(t)\big] \label{eq:fma_a-lin}\\
\od{a_{n,1}}{t} &=  \lambda_n(h_0) a_{n,1} + \od{h_1}{t}c_{n0}(h_0)   \label{eq:fma_b-lin},
\end{align}
\end{subequations}
Note that in Eq.~\eqref{eq:fma_b-lin} only the term $c_{n0}(h_0)$ survives because $dh_1/dt = \mathcal{O}(\varepsilon)$ and $\sum_m a_m c_{nm} = c_{n0} + \mathcal{O}(\varepsilon)$.
Applying the Fourier transform, $\tilde{f}(\omega)=\int_{-\infty}^\infty e^{-i\omega t}f(t)\,dt$, to Eq.~\eqref{eq:mode-approx-lin}, yields
\begin{align}
  \label{eq:fourier-spec}
  \tilde{A_1}(\omega)=F_0'(h_0)\tilde{h}_1+\sum_{n=1}^\infty\lrrund{F_n(h_0)\widetilde{a_{n,1}}(\omega)+F_n^*(h_0)\widetilde{a_{n,1}^*}(\omega)},
\end{align}
  where $\widetilde{a_{n,1}}$ and $\widetilde{a_{n,1}^*}$ satisfy
  \begin{align}
  i\omega\widetilde{a_{n,1}}(\omega)&=\lambda_n(h_0)\widetilde{a_{n,1}}(\omega)+i\omega c_{n0}(h_0)\tilde{h}_1(\omega)\\
  i\omega\widetilde{a_{n,1}^*}(\omega)&=\lambda_n^*(h_0)\widetilde{a_{n,1}^*}(\omega)+i\omega c_{n0}^*(h_0)\tilde{h}_1(\omega).
\end{align}
Eliminating $\widetilde{a_{n,1}}$ and $\widetilde{a_{n,1}^*}$ yields a relation between the population activity and the input in frequency space:
\begin{equation}
  \label{eq:A-h1}
\tilde{A}_1(\omega) = \chi_{\infty,h}(\omega) \tilde{h}_1(\omega)
\end{equation}
where we introduced the linear response function $\chi_{\infty,h}(\omega)$ (with respect to weak modulation in $h(t)$),
\begin{equation}
  \label{eq:mode-lin-respon}
  \chi_{\infty,h}(\omega)=F_0'(h_0)  + i\omega \sum_{n=1}^\infty \left( \frac{F_n(h_0) c_{n0}(h_0)}{i\omega -  \lambda_n(h_0)} +\frac{F_n^*(h_0)c_{n0}^*(h_0)}{i\omega - \lambda^*_n(h_0)}\right).
\end{equation}
Truncating the infinite series in Eq.~\eqref{eq:mode-approx-lin} and ~\eqref{eq:mode-lin-respon} at $n=1$, yields the first-order approximation of the linear response function
\begin{equation}
  \label{eq:mode-lin-respon}
  \chi_{1,h}(\omega)=F_0'(h_0)  + i\omega \left( \frac{F_1(h_0) c_{10}(h_0)}{i\omega -  \lambda_1(h_0)} +\frac{F_1^*(h_0)c_{10}^*(h_0)}{i\omega - \lambda^*_1(h_0)}\right),
\end{equation}
see also Eqs.~\eqref{eq:first-mode-approx-lin} and ~\eqref{eq:lin-respon} in the main text.

According to Eq.~\eqref{eq:h_int}, the input  $h(t)=h_0+\epsilon h_1(t)$ consists of external and recurrent inputs. Specifically, we consider $h_\text{ext}(t)=[\kappa*(I_0+\varepsilon\mu_1)](t)$; hence,
\begin{align}
  h_0&= \bar\kappa I_0 + J \bar\epsilon A_0,\\
  \label{eq:app_c1}
h_1(t)&= (\kappa \ast \mu_1)(t) + J (\epsilon \ast A_1)(t),
\end{align}
where $\bar\kappa=\int_0^\infty\kappa(t)\,dt$, $\bar\epsilon=\int_0^\infty\epsilon(t)\,dt$.
In Fourier space, Eq.~\eqref{eq:app_c1} becomes
\begin{equation}\label{eq:appc1_h1}
\tilde{h}_1(\omega) = \tilde\kappa(\omega) \tilde{\mu}_1(\omega) + J \tilde{\epsilon}(\omega) \tilde{A}_1(\omega).
\end{equation}
Together with Eq.~\eqref{eq:A-h1}, we find
\begin{equation}
  \label{eq:A-mu1}
  \tilde{A}_1(\omega)=\chi_\mu(\omega)\tilde{\mu}_1(\omega),
\end{equation}
where
\begin{equation}
  \label{eq:linresp-mu}
  \chi_\mu(\omega)=\frac{\tilde{\kappa}(\omega)\chi_h(\omega)}{1-J\tilde{\epsilon}(\omega)\chi_h(\omega)}
\end{equation}
is the linear response function with respect to changes $\mu_1(t)$ in the input current $I(t) = I_0 + \mu_1(t) $.
The linear response function $ \chi_\mu(\omega)$ is fully determined by the filter functions and the linear response function $\chi_h(\omega)$ with respect to changes in $h(t)$.

For the numerical examples in the main text, we used the first-order dynamics $\tau_h \dot h = - h + I(t)$ of the input potential, see Eq.~\eqref{eq:h-first-ord}, corresponding to an exponential filter $\kappa(t)$ with Fourier transform 
\begin{equation}\label{eq:appc1_kappa}
\tilde{\kappa}(\omega) = \frac{1}{1 + i \omega \tau_h}.
\end{equation}


\subsection{Poisson model with absolute refractoriness}

To compare the linear response of the first-order approximation with (i) the exact linear response and (ii) with the linear response predicted by a classical firing rate model, we considered in Sec.~\ref{sec:popul-dynam-resp} and \ref{sec:popul-dynam-inter} the Poisson model with absolute refractory period $\Delta$  and input-dependent rate parameter $\nu(h)$ (PAR model). This model provides closed-form analytical expressions for these reference cases.

\subsubsection{Exact linear response based on integral equation}
\label{sec:linearizeddyn-integr-par}

To obtain the exact linear response, we linearize the refractory density equation about the stationary solution $A_0$. In the PAR model, this refractory density equation reduces to the integral equation \cite{GerKis14}
\begin{equation}
\label{eq:Pois_integraleq}
A(t) = \nu\big( h(t)\big)  \big[ 1-  \int_{t-\Delta}^t A(s)\d s\big]. 
\end{equation}
Inserting $A(t)=A_0+\varepsilon A_1(t)$ and $h(t)=h_0+\varepsilon h_1(t)$ into Eq.~\eqref{eq:Pois_integraleq} and expanding  to first order yields
\if \draftmode1
\begin{multline}
  A_0+\varepsilon A_1(t)=[\nu(h_0)+\varepsilon \nu'(h_0)h_1(t)]\\
  \times\lrrund{1-\int_{t-\Delta}^t[A_0+\varepsilon A_1(t')]\,dt'}.
\end{multline}
\else
\begin{equation}
  A_0+\varepsilon A_1(t)=[\nu(h_0)+\varepsilon \nu'(h_0)h_1(t)]\lrrund{1-\int_{t-\Delta}^t[A_0+\varepsilon A_1(t')]\,dt'}.
\end{equation}
\fi
Collecting first-order terms, we obtain
\begin{align}
  A_1(t)&=-\nu(h_0)\int_{t-\Delta}^tA_1(t')\,dt'\nonumber\\
  &\quad+\nu'(h_0)h_1(t)\lrrund{1-\int_{t-\Delta}^tA_0\,dt'}\\
        &=-\nu(h_0)\int_{0}^\Delta A_1(t-s)\,ds\nonumber\\
  &\quad+\nu'(h_0)h_1(t)\lrrund{1-\Delta A_0}.
\end{align}
Applying the Fourier transform to both sides of the last equation and rearranging terms, we have
\begin{align}
  \tilde{A}_1(\omega)&=-\nu(h_0)\int_{0}^\Delta \int_{-\infty}^\infty e^{-i\omega t}A_1(t-s)\,dtds\nonumber\\
  &\quad+\nu'(h_0)\tilde{h}_1(\omega)\lrrund{1-\Delta A_0},\\
                     &=-\nu(h_0)\tilde{A}_1(\omega)\frac{1-e^{-i\omega\Delta}}{i\omega}\nonumber\\
  &\quad+\nu'(h_0)\tilde{h}_1(\omega)\lrrund{1-\Delta A_0}.
\end{align}
Thus,
\begin{equation}
  \tilde{A}_1(\omega)=\chi_h(\omega)\tilde{h}_1(\omega)
\end{equation}
with the exact linear response function
\begin{equation}
  \tilde{\chi}_h(\omega)=\frac{\nu'(h_0)\lrrund{1-\Delta A_0}}{1+\nu(h_0)\frac{1-e^{-i\omega\Delta}}{i\omega}}.
\end{equation}

\subsubsection{Linear response for low-dimensional firing rate model (first-order approximation)}

In the first-order approximation, the linear response function $\chi_h(\omega)$ is given by $\chi_{1,h}(\omega)$ in Eq.~\eqref{eq:mode-lin-respon} with
\begin{equation}
F_0'(h_0) = \frac{\nu'(h_0) \big(1-\Delta A_0\big)}{1+\Delta \nu(h_0)}  
\end{equation}
and
\begin{equation}
  c_{n0}(h_0)=\frac{\nu'(h_0)}{[\nu(h_0)+\lambda_1(h_0)][1+\Delta\nu(h_0)]}
\end{equation}
(cf. Eqs.~\eqref{eq:F0-par} and ~\eqref{eq:Pois_cnm}). Furthermore, $\lambda_1(h_0)$ and $F_1(h_0)$ are given by Eqs.~\eqref{eq:Pois_lambdan} and \eqref{eq:Fn-par}.

\subsection{Stability analysis of asynchronous activity}
\label{sec:stab-analys-asynchr}

We can exploit the linear response function $\chi_h$ to determine the stability of the stationary solution and also find oscillatory instabilities of the population activity, giving rise to collective oscillations and synchrony \cite{BruWan03,GerKis14}.
We reconsider the linear system Eqs.~(\ref{eq:A-h1}~\!\&\!~\ref{eq:appc1_h1})
about a stationary solution $A(t)=A_0$ with constant external input $I(t) =I_0$, i.e.~$\mu_1(t)=0$,
that is, 
\begin{equation}\label{eq:c2_A1}
\begin{aligned}
\tilde{A}_1(\omega) &= \chi_h(\omega) \tilde{h}_1(\omega)\\
\tilde{h}_1(\omega) &= J \tilde\varepsilon(\omega) \tilde{A}_1(\omega).
\end{aligned}
\end{equation}
Focusing on solutions of the form
\begin{equation}\label{eq:A_1_sol_form}
\begin{aligned}
A_1(t) &= \hat{A}_1(\lambda) \mathrm{e}^{\lambda t} + \hat{A}^*_1(\lambda) \mathrm{e}^{\lambda^* t}, \\
h_1(t) &= \hat{h}_1(\lambda) \mathrm{e}^{\lambda t} + \hat{h}^*_1(\lambda) \mathrm{e}^{\lambda^* t},
\end{aligned}
\end{equation}
we can plug Eq.~\eqref{eq:A_1_sol_form} into Eq.~\eqref{eq:c2_A1}, which leads to
\begin{equation}
\begin{aligned}
\hat{A}_1(\lambda) &= {\chi}_h(-i\lambda) \hat{h}_1(\lambda),\\
\hat{h}_1(\lambda) &= J \tilde{\epsilon}(-i\lambda) \hat{A}_1(\lambda).
\end{aligned}
\end{equation}
Taken together, we obtain the \emph{characteristic equation} 
\begin{equation}\label{eq:appc2_chareq}
1 = J\tilde{\epsilon}(-i\lambda) {\chi}_h(-i\lambda) := \chi_c(-i\lambda) 
\end{equation}
with complex-valued argument $\lambda \in \mathbb{C}$.
The stationary solution is stable (unstable) if $\mathrm{Re}\,(\lambda)<0$ ($\mathrm{Re}\,(\lambda)>0$).
Moreover, we find an oscillatory instability for $\lambda = i \omega$, $\omega\in\mathbb{R}$, $\omega>0$, and the characteristic equation~\eqref{eq:appc2_chareq} becomes
\begin{equation}
\chi_c(\omega) = |\chi_c(\omega) | \mathrm{e}^{i\arg(\chi_c(\omega)) } =  1.
\end{equation}
That is, at an oscillatory instability, the two conditions
\begin{equation}\label{eq:appc2_phaseampl1}
|\chi_c(\omega)| = 1 \quad \text{and} \quad \arg\big( \chi_c(\omega) \big) = 0 \mod 2\pi
\end{equation}
have to be fulfilled simultaneously such that collective oscillations can emerge at a critical coupling strength $J_c$ and with a critical frequency $\omega_c$.
To be more precise, we write the complex-valued functions $\tilde{\epsilon}(\omega)$ and ${\chi}_h(\omega)$ in polar form:
\begin{align*}
\tilde{\epsilon}(\omega) &= \big| \tilde{\epsilon}(\omega) \big| \, \mathrm{e}^{i\Phi_\epsilon(\omega)},\\
{\chi}_h(\omega) &= \big| {\chi}_h(\omega) \big| \, \mathrm{e}^{i\Phi_\chi(\omega)}.
\end{align*}
Inserting them into Eq.~\eqref{eq:appc2_chareq} results in a phase condition and an amplitude condition akin to Eq.~\eqref{eq:appc2_phaseampl1}.
The \emph{phase condition} reads:
\begin{align}\label{eq:appc2_phasecond}
\Phi_\epsilon(\omega) + \Phi_\chi(\omega) = \begin{cases} 2 k \pi, \quad &J>0 \text{ (exc. population)},\\
(2 k+1) \pi, &J<0 \text{ (inh. population)},\end{cases}
\end{align}
and yields the critical frequency $\omega_{c,k}$ for $k=1,2,\dots$ in a population with excitatory ($J>0$) or inhibitory ($J<0$) coupling, respectively.
We can use $\omega_{c,k}$ to determine the critical coupling $J_{c,k}$ by solving the \emph{amplitude condition}
\begin{equation}\label{eq:appc2_amplcond}
|J_{c,k}| = \frac{1}{| \tilde{\epsilon}(\omega_{c,k})| \cdot | {\chi}_h(\omega_{c,k}) |}.
\end{equation}
Note, however, that in general the phase and amplitude conditions Eqs.~(\ref{eq:appc2_phasecond}\,\&\,\ref{eq:appc2_amplcond}) have to be solved simultaneously as the linear response $\chi_h$ depends implicitly on the coupling strength $J$ through $A_0=F_0(I_0 + JA_0)$, where $F_0$ is the f-I curve, or transfer function.

As the approach above only requires knowledge about the linear response function $\chi_h$, it can readily be applied for the three different models of Poisson neurons with absolute refractoriness. 
In particular, it can be used for distinguishing whether the first-order approximation can indeed capture the onset of oscillations across parameter space.




 \makeatletter
\setcounter{equation}{0}
\renewcommand{\theequation}{C\arabic{equation}}
 \makeatother
  
\section{On the equivalence of Fokker-Planck and refractory density approaches}
\label{sec:equiv-fokk-planck}
Dealing with one-dimensional integrate-and-fire dynamics, a common approach to describe the population activity of such neurons dwells on the Fokker-Planck formalism.
Eigenfunction expansions of the associated Fokker-Planck operator have been performed by \cite{KniMan96,Kni00,MatGiu02,SchOst13,Mat16_arxiv,AugLad17}.
As described above, the refractory density approach represents a more general alternative.
Yet, as the latter is based on the ISI distribution $P(\tau)$ that results from a first-passage time problem, it is not obvious that a spectral decomposition of the refractory density operator will result in the same set of eigenvalues and hence the same (low-dimensional) firing rate model.
In particular, as the Fokker-Planck operators corresponding to the firing rate dynamics and to the first-passage time statistics have different boundary conditions, the respective spectra will in general be different.
Here we show, however, that the eigenvalue formula of the refractory density approach imposes a certain relation between the ISI distribution and the spectrum, which yields identical boundary conditions as in the Fokker-Planck approach for the firing rate dynamics.
Conclusively, we can prove that the associated spectra of Fokker-Planck and refractory density approaches coincide and thus lead to same firing rate model.
As our result holds for an arbitrary absolute refractory period after firing, it opens the possibility to go beyond the work of Mattia and co-workers \cite{MatGiu02} and to derive eigenfunction expansions for integrate-and-fire dynamics with refractoriness.

\subsection{Fokker-Planck approach}
To begin, we briefly revisit the Fokker-Planck approach for the one-dimensional integrate-and-fire dynamics 
\begin{equation}
  \label{eq:langevin}
\dot{v} =  \mu(v) + \sigma(v) \xi(t),
\end{equation}
where a possible membrane time constant $\tau_m$ is incorporated in the drift and diffusion functions, $\mu(v)$ and $\sigma(v)$, respectively.
The quantity $\xi(t)$ denotes a Gaussian white noise term and Eq.~\eqref{eq:langevin} is interpreted in the Ito sense.
In general, $\mu$ and $\sigma$ may be subjected to time-dependent input, describing either some external  current or recurrent network activity.
To determine the eigenvalues, we consider again the time-homogeneous case, in which $\mu=\mu(v)$ and $\sigma=\sigma(v)$ do not possess an explicit time-dependence but allow for an explicit dependence on the membrane potential $v$.
Whenever $v$ hits an upper threshold $\vth$, an action potential is elicited and after some time $\Delta$, the absolute refractory period, the membrane potential is reset to some lower value $\vreset$.
We may further impose a reflecting boundary at some membrane potential $v_0\geq - \infty$, beyond which no values of $v$ are allowed.

The time evolution of the corresponding probability density $q(v,t)$ is governed by the Fokker-Planck equation \cite{Ris84,van92}
\begin{align}\label{eq:FP_equation}
  \frac{\partial}{\partial t} q(v,t) &= \Big[ -\frac{\partial}{\partial v} \mu(v) + \frac{1}{2} \frac{\partial^2}{\partial v^2}  \sigma^2(v) \Big] q(v,t)\\
  &\equiv \op{L}_{v} q(v,t). \nonumber 
\end{align}
The associated flux is denoted as $J(v,t) = \big[ \mu(v) - \tfrac{1}{2} \partial_v \sigma^2(v)\big] q(v,t)$ and the boundary conditions are
\begin{subequations}\label{eq:app_boundcondall}
\begin{align}
q(\vth,t) &= 0 \label{eq:app_boundcond1}\\
J(v_0,t) &= 0  \label{eq:app_boundcond2}\\
J(\vth,t-\Delta) &= J(\vreset^+,t) - J(\vreset^-,t).  \label{eq:app_boundcond3}
\end{align}
\end{subequations}
The first condition Eq.~\eqref{eq:app_boundcond1} represents an absorbing boundary at threshold, the second Eq.~\eqref{eq:app_boundcond2} a reflecting boundary at $v=v_0$.
The third condition Eq.~\eqref{eq:app_boundcond3} guarantees the conservation of flux by taking into account the reset following the firing of an action potential; the abbreviation
$f(v^\pm) = \lim_{ \varepsilon \to 0} f(v + \pm \varepsilon)$ indicates whether we perform the limit from above or below.

The Fokker-Planck operator $\op{L}_{v}$ has a set of eigenfunctions $\tilde\phi_n(v)$ associated with eigenvalues $\lambda_n$,
\begin{equation}\label{eq:app_eqC4}
\op{L}_{v} \tilde\phi_n(v) = \tilde\lambda_n \tilde\phi_n(v).
\end{equation}
With the inner product defined as $\langle \tilde\phi | \tilde\psi \rangle =\int\tilde\psi(v)\tilde\phi(v)dv$, we can define the adjoint operator $\op{L}_{v}^+$ having eigenfunctions $\tilde\psi_m$ associated with the same eigenvalues $\tilde\lambda_m$.
Assuming that $\lbrace\tilde\phi_n\rbrace_{n\in\mathbb{Z}}$ is a complete set of eigenfunctions, the definition of the adjoint operator implies that eigenfunctions with different eigenvalues are  orthogonal to each other.
For the sake of completeness, Eq.~\eqref{eq:app_eqC4} reads
\begin{equation}\label{eq:FP_2ndorderODE}
\tilde\lambda_n \tilde\phi_n(v) = \big[ -\partial_v \mu(v) + \tfrac{1}{2} \partial_v^2 \sigma^2(v) \big] \tilde\phi_n(v)
\end{equation}
for $v\in (v_0,\vth)$,
with $\tilde\phi_n(v)$ respecting the boundary conditions Eq.~\eqref{eq:app_boundcondall}.
In practice, to solve for the eigenvalues $\tilde\lambda_n$, one determines general solution formulae for $\tilde\phi_n$ as well as for the eigenfunctions $\tilde\psi_m$ of $\op{L}_{v}^+$ subject to an appropriate normalization $\langle \tilde\phi_n | \tilde\psi_m \rangle = \delta_{nm}$ and satisfying a corresponding set of boundary conditions.
In principle, however, it suffices to impose the boundary conditions Eq.~\eqref{eq:app_boundcondall} on the general solution for $\tilde\phi_n$ to find the associated $\tilde\lambda_n$.
As has been shown in the literature \cite{Kni00, MatGiu02}, the following four remarks hold for the eigenvalues $\tilde\lambda_n$ of the Fokker-Planck operator $\op{L}_{v}$:
First, the $\tilde\lambda_n$ appear in general in complex conjugate pairs due to the re-entering flux condition
at reset. Without that condition, $\op{L}_{v}$ would admit only real-valued eigenvalues \cite{RicSat88,Ost11}.
Second, the $\tilde\lambda_n$ have negative real part for $n \neq 0$.
Third, there always exists a null eigenvalue $\tilde\lambda_0=0$ corresponding to the stationary solution $q_0(v)$ of Eq.~\eqref{eq:FP_equation}.
And fourth, due to the bi-orthogonality, the associated (adjoint) eigenfunction $\tilde\psi_0(v) \equiv 1$ is constant.

\subsection{Refractory density approach}
As introduced in detail in the main text, the refractory density approach dwells on the hazard rate describing the firing probability of the neurons.
The connection between hazard rate $\rho(\tau,h)$, survival probability $S(\tau)$, and ISI distribution $P(\tau)$ allows for reformulating the integrate-and-fire dynamics in terms of the refractory density $p(\tau,t)$ instead of the membrane potential distribution $q(v,t)$.
The master equation governing its time evolution is the transport equation
$$\partial_t p(\tau,t) = \big[ - \partial_\tau - \rho(\tau,h) \big] p(\tau,t) \equiv \op{L}_\tau p(\tau,t).$$
The eigenfunctions ${\phi}_n(\tau)$ of the refractory density operator $\op{L}_\tau$, satisfying $\op{L}_\tau{\phi}_n=\lambda_n {\phi}_n$, have been found as
\begin{equation}\label{eq:L_eigenfunction}
{\phi}_n(\tau) = {\phi}_n(0) \mathrm{e}^{-\lambda_n \tau} S(\tau)
\end{equation} 
and the eigenvalues $\lambda_n$ are the roots of 
\begin{equation}\label{eq:PLeigenvalueformula}
P_L(\lambda_n) - 1 =  0 ,
\end{equation}
where the subscript $L$ indicates the Laplace transform of the ISI density $P_L(s) = \int_0^\infty  \mathrm{e}^{-s \tau} P(\tau) d\tau$.
One may already anticipate the complex nature of the eigenvalues from Eq.~\eqref{eq:PLeigenvalueformula} as well as that they have real negative part and that there exists one null eigenvalue $\lambda_0=0$, very similar to the remarks above about the spectrum of the Fokker-Planck operator $\op{L}_{v}$.
Note, however, that the eigenvalues $\lambda_n$ of the refractory density operator $\op{L}_\tau$ are in general different from those of the $\op{L}_{v}$.
In the following, we will prove that the respective spectra of $\op{L}_\tau$ and of $\op{L}_{v}$ indeed coincide, i.e. $\tilde\lambda_n=\lambda_n$.

To do so, we first focus on deriving the ISI distribution $P(\tau)$ according to the definition
\begin{equation}\label{eq:P_ISIdistribution}
P(\tau) =  J(\vth,\tau) = -\tfrac{1}{2} \partial_v \sigma^2(v) q(v,\tau) \big|_{v = \vth}
\end{equation}
where $q(v,\tau)$ is the transition probability density function and follows the above-introduced Fokker-Planck equation Eq.~\eqref{eq:FP_equation} with the boundary conditions Eqs.~(\ref{eq:app_boundcond1}~\!\&~\!\ref{eq:app_boundcond2}).
Instead of the flux condition Eq.~\eqref{eq:app_boundcond3}, however, $q(v,\tau)$ is now subject to the initial condition 
\begin{equation}\label{eq:initialcondition}
q(v,\Delta) = \delta(v-\vreset )
\end{equation}
with $\Delta\geq 0$ the absolute refractory period; in comparison to the main text, we here abbreviated $q(v,\tau) = q(v,\tau | \vreset )$.
Since the reset mechanism as implemented by the flux condition Eq.~\eqref{eq:app_boundcond3} is ignored, the spectrum of the Fokker-Planck operator for the transition probability density function is different from that for the rate dynamics above.
However, in combining the initial condition Eq.~\eqref{eq:initialcondition} with the eigenvalue formula Eq.~\eqref{eq:PLeigenvalueformula}, the reset mechanism will be reinstated and thus enforces the spectrum to coincide with that of the rate dynamics.

The proof builds on the fact that the Laplace transform $q_L(v,s)$ of the transition probability density function $q(v,\tau)$ evaluated at the roots $s=\lambda_n$ of the eigenvalue formula Eq.~\eqref{eq:PLeigenvalueformula} satisfies the same second-order ordinary differential equation Eq.~\eqref{eq:FP_2ndorderODE} with boundary conditions Eq.~\eqref{eq:app_boundcondall} as do the eigenfunctions $\tilde\phi_n(v)$ of the Fokker-Planck operator $\op{L}_{v}$ with associated eigenvalues $\tilde\lambda_n$.
Existence and uniqueness of solutions for the boundary value problem Eqs.~(\ref{eq:FP_2ndorderODE}~\!,~\!\ref{eq:app_boundcond1}~\!\&~\!\ref{eq:app_boundcond2}) hold as long as $\mu(v)$ and $\sigma(v)$ are continuous (as assumed) on the open interval $(v_0,\vth)$ and the third condition Eq.~\eqref{eq:app_boundcond3} vis-\`a-vis the eigenvalue formula Eq.~\eqref{eq:PLeigenvalueformula} together with the initial condition Eq.~\eqref{eq:initialcondition} constraints the associated eigenvalue $\lambda_n$ as a constant parameter.

To show the above, we consider the Laplace transform of the transition probability density function $q(v,\tau)$,
$$
q_L(v,s) = \int_{-\infty}^\infty \mathrm{e}^{-s\tau} q(v,\tau) d\tau =\int_{\Delta}^\infty  \mathrm{e}^{-s\tau} q(v,\tau) d\tau,
$$
where we used that $q(v,\tau)$ has support on the interval $(\Delta,\infty)$.
$q_L(v,s)$ follows the (Laplace transformed) dynamics
\begin{equation}\label{eq:laplacedynamics1}
-\mathrm{e}^{-\Delta s} \delta(v-\vreset ) + sq_L(v,s) = \big[ - \partial_v \mu(v) + \tfrac{1}{2} \partial_v^2 \sigma^2(v) \big] q_L(v,s)
\end{equation}
with boundary conditions
\begin{align}\label{eq:P_boundaryconditions}
q_L(\vth,s) = 0 \quad \text{ and } \quad J_L(v_0,s) = 0
\end{align}
where $J_L$ is the Laplace transform of the flux $J(v,\tau)$ associated with Eq.~\eqref{eq:FP_equation}.
We can rewrite the $\delta$-function in the dynamics Eq.~\eqref{eq:laplacedynamics1} in form of an additional jump condition
\begin{equation}
\label{eq:laplace_jumpcondition1}
-\mathrm{e}^{-\Delta s} = \frac{1}{2} \Big[ \partial_v \sigma^2(\vreset ) q(\vreset^+,s) - \partial_v \sigma^2(\vreset ) q(\vreset^-,s) \Big] 
\end{equation}
where we integrated Eq.~\eqref{eq:laplacedynamics1} over an $\varepsilon$-neighborhood of $\vreset $ and took the limit $\varepsilon \to 0$ whilst capitalizing on the continuity of $q_L(v,s), \mu(v)$, and $\sigma(v)$ on $(v_0,\vth)$.
We can then use the absorbing boundary condition at threshold to reformulate the jump condition Eq.~\eqref{eq:laplace_jumpcondition1} as
\begin{equation}\label{eq:laplace_jumpcondition2}
\mathrm{e}^{-\Delta s}  = J_L(\vreset^+,s) - J_L(\vreset^-,s).
\end{equation}
Albeit similar in nature, Eq.~\eqref{eq:laplace_jumpcondition2} is still distinct from the flux condition Eq.~\eqref{eq:app_boundcond3}.
At this point, however, we have not exploited the eigenvalue formula Eq.~\eqref{eq:PLeigenvalueformula}, yet, which we will do next.
Recall that the Laplace transform $P_L(s)$ of the ISI distribution evaluated at the eigenvalues $s=\lambda_n$ equals unity. 
The left-hand side of Eq.~\eqref{eq:laplace_jumpcondition2} can thus be written as
\begin{equation}\label{eq:laplace_jumpcondition3}
\mathrm{e}^{-\Delta s} = 1 \cdot \mathrm{e}^{-\Delta s} = P_L(s) \mathrm{e}^{-\Delta s} \big|_{s=\lambda_n} = J_L(\vth,\lambda_n)\mathrm{e}^{-\Delta \lambda_n},
\end{equation}
where we used Eq.~\eqref{eq:P_ISIdistribution} for the last equality.
Now, combining Eqs.~(\ref{eq:laplace_jumpcondition2}~\!\&~\!\ref{eq:laplace_jumpcondition3}) and applying the inverse Fourier transform (with respect to $s=\lambda_n$), we arrive at
\begin{equation}
J(\vth, t - \Delta) = J(\vreset^+,t) - J(\vreset^-,t),
\end{equation}
which coincides with the flux condition Eq.~\eqref{eq:app_boundcond3}.
In total, we have that $q_L(v,\lambda_n)$ satisfies the boundary value problem
\begin{align}
\lambda_n q_L(v,\lambda_n) = \big[ - \partial_v \mu(v) + \tfrac{1}{2} \partial_v^2 \sigma^2(v) \big] q_L(v,\lambda_n) 
\end{align}
on $v \in (v_0,\vth)$ with boundary conditions Eq.~\eqref{eq:P_boundaryconditions} and flux condition Eq.~\eqref{eq:app_boundcond3}.
Hence, the $q_L(v,\lambda_n)$ are identical (upon re-ordering if necessary) with the eigenfunctions $\tilde\phi_n(v)$ of the Fokker-Planck operator $\op{L}_{v}$ associated with the eigenvalues $\tilde\lambda_n$.
In particular, we can conclude that $\tilde{\lambda}_n=\lambda_n$ and that the spectrum of the refractory density operator $\op{L}_\tau$ lies in the left complex half plane with one null eigenvalue $\lambda_0=0$.

What is more, we can express the eigenfunctions ${\phi}_n(\tau)$ of the refractory density operator $\op{L}_\tau$ in terms of the eigenfunctions $\tilde\phi_n(v)$ of the Fokker-Planck operator $\op{L}_{v}$.
Using the definition of the survival probability $S(\tau) = 1- \int_0^\tau P(s)ds$ together with Eq.~\eqref{eq:P_ISIdistribution}, we have
\begin{equation}\label{S_eigenfunc}
S(\tau) = 1 + \frac{1}{2}\int_0^\tau \partial_v \sigma^2(v)  \widehat{\phi_n(v)}(s) \big|_{v=\vth} ds,
\end{equation} 
where 
$$\widehat{\phi_n(v)}(s) = \mathcal{F}^{-1} \big( \tilde\phi_n(v) \big) (s) $$
is the inverse Fourier transform of the eigenfunction $\tilde\phi_n(v)$ with respect to its associated eigenvalue $\lambda_n$.
Eq.~\eqref{S_eigenfunc} can then be inserted in the formula Eq.~\eqref{eq:L_eigenfunction} for the eigenfunction ${\phi}_n(\tau)$ of the refractory density operator $\op{L}_\tau$.
Note, however, that only the direction $\tilde\phi_n(v) \mapsto \phi_n(\tau)$ works, but not the other way around.

In conclusion, we have proven that Fokker-Planck operators for the rate dynamics (with flux condition but without initial condition) and for the first-passage time statistics (no flux, but initial condition) yield the same set of eigenvalues for one-dimensional integrate-and-fire dynamics if and only if the eigenvalue formula Eq.~\eqref{eq:PLeigenvalueformula} of the refractory density approach is invoked.


 \makeatletter
\setcounter{equation}{0}
\renewcommand{\theequation}{D\arabic{equation}}
 \makeatother
 
\section{Known statistics of perfect and leaky integrate-and-fire models}
\label{sec:known-stat-perf}

In Sec.~\ref{subsec:if-models}, we considered the perfect (PIF) and leaky integrate-and-fire (LIF) models.
The ISI density of the PIF neuron model has the form of an inverse Gaussian distribution,
\begin{equation}
\label{eq:inversegaussian}
P(\tau,h)=\frac{\vth }{\sqrt{4\pi D\tau^3}}\exp\left(-\frac{(\mu(h)\tau-\vth )^2}{4D\tau}\right),
\end{equation}
and its rate $r$ and CV are given by 
$$r=\frac{\mu(h)}{\vth }, \qquad \cv=\displaystyle{\sqrt{\frac{2D}{\mu(h)\vth }}}.$$

The rate $r$ and the CV of the LIF model (with rescaled membrane potentials such that $\vreset=0$ and $\vth=1$) are given by (e.g. \cite{LinLSG02})
\begin{equation}
r = \frac{1}{\langle \tau \rangle} \quad \text{and} \quad C_V = \frac{\sqrt{\langle \Delta\tau^2 \rangle}}{\langle \tau \rangle}  
\end{equation}
with
\begin{equation}
\langle \tau \rangle = \sqrt{\pi} \int_{(\mu-1)/\sqrt{2D}}^{(\mu)/\sqrt{2D}} \mathrm{e}^{y^2} \mathrm{erfc}(y) dy  
\end{equation}
and
\if \draftmode1
\begin{equation}
\langle \Delta\tau^2 \rangle = 2\pi \int_{(\mu-1)/\sqrt{2D}}^{\infty}dy\, \mathrm{e}^{y^2} \big[\mathrm{erfc}(y)\big]^2 \int_{(\mu-1)/\sqrt{2D}}^{y}dz\, \mathrm{e}^{z^2} \Theta\Big( \frac{\mu}{\sqrt{2D}} - z \Big),
\end{equation}
\else
\begin{align}
\langle \Delta\tau^2 \rangle = &2\pi \int_{(\mu-1)/\sqrt{2D}}^{\infty}dy\, \mathrm{e}^{y^2} \big[\mathrm{erfc}(y)\big]^2 \nonumber\\
&\times \int_{(\mu-1)/\sqrt{2D}}^{y}dz \mathrm{e}^{z^2} \Theta\Big( \frac{\mu}{\sqrt{2D}} - z \Big),
\end{align}
\fi
where $\mathrm{erfc}(y)$ is the complementary error function and $\Theta(x)$ denotes the Heaviside function.


 \makeatletter
\setcounter{equation}{0}
\renewcommand{\theequation}{E\arabic{equation}}
 \makeatother
 
\section{Alternative derivation of the characteristic equation \eqref{eq:condition2}}
\label{sec:alt-deriv}

To begin, we rewrite the refractory density equation (\ref{eq:refr-dens}~\!\&\!~\ref{eq:popact-integr}) for the refractory density equation $p(\tau,t)$ as follows:
\begin{equation}
\partial_t p(\tau,t) = - \partial_\tau p - \rho(\tau) p  + A(t) \delta(\tau), \quad t,\tau \ge 0,
\end{equation}
where $A(t) = \int_0^\infty \rho(\tau) p(\tau,t) d\tau = p(0,t)$ and $\delta(\tau)$ is the delta function. Here and in the following, we will omit the explicit dependence on the momentary parameter $h$ for conciseness, but remark that an extension for $h$ being time-varying, and/or multi-dimensional can readily be obtained.
The Laplace transform
$\hat p(\tau,s) = \int_0^\infty \mathrm{e}^{-st} p(\tau,t)dt$ of $p(\tau,t)$ with respect to the time argument $t$ can be found (by partial integration) to follow the Laplace transformed dynamics
\begin{equation}
- p(\tau,0) + s \hat p(\tau,s) = \partial_\tau \hat p - \rho(\tau) \hat p + \hat A(s) \delta(\tau)
\end{equation}
with $\hat A(s)$ the Laplace transform of $A$ and provided that $\lim_{t\to\infty} \mathrm{e}^{-st}p(\tau,t) = 0$.
Recall that $p(\tau,0)$ is the initial distribution density of the neurons' refractory states.
Rearranging terms, we thus have the inhomogeneous ordinary differential equation
\begin{equation}\label{eq:appE_phat_dyn}
\big[ \partial_\tau + s + \rho(\tau) \big] \hat p(\tau,s) =  p(\tau,0) +  \hat A(s) \delta(\tau)
\end{equation}
We next define $\hat q(\tau,s)$ such that 
\begin{align}
\hat p(\tau,s) &= \exp\Big( -s\tau - \int_0^\tau \rho(x)dx \Big) \hat q(\tau,s)\nonumber\\ &=\mathrm{e}^{-s\tau}S(\tau) \hat q(\tau,s).\label{eq:appE_qhat_def}
\end{align}
The dynamics Eq.~\eqref{eq:appE_phat_dyn} in the new variable $\hat q(\tau,s)$ then become
\begin{align}
\partial_\tau \hat q(\tau,s) &= \big[  p(\tau,0) +  \hat A(s) \delta(\tau) \big] \frac{\mathrm{e}^{s\tau}}{S(\tau)}\nonumber\\
&= p(\tau,0) \frac{\mathrm{e}^{s\tau}}{S(\tau)} +  \hat A(s) \delta(\tau) ,\label{eq:appE_qhat_dyn}
\end{align}
since $\mathrm{e}^{s\tau}/S(\tau)\big|_{\tau=0} = 1$,
and with initial condition $\hat q(0,s) = \hat A(s)$.
The general solution of Eq.~\eqref{eq:appE_qhat_dyn} thus reads
\begin{equation}\label{eq:appE_qhat_sol}
\hat q(\tau,s) = \int_0^\tau p(x,0) \frac{\mathrm{e}^{sx}}{S(x)}dx + \hat q(s,0).
\end{equation}
With $P(\tau)/S(\tau)=\rho(\tau)$ and Eq.~\eqref{eq:appE_qhat_def}, we can reformulate the Laplace transform of $A(t)$ as
\begin{equation}
\hat A(s) = \int_0^\infty \mathrm{e}^{-s \tau} P(\tau) \hat q(\tau,s) d\tau \ .
\end{equation}
Recalling the initial condition $\hat A(s) = \hat q(0,s)$, we now multiply Eq.~\eqref{eq:appE_qhat_sol} by $\mathrm{e}^{-s \tau} P(\tau)$ and integrate the product with respect to $\tau$, which yields
\begin{align}
  \hat q(0,s) &= \hat A(s)\nonumber\\
              &= \int_0^\infty \mathrm{e}^{-s \tau} P(\tau) \int_0^\tau p(x,0) \frac{\mathrm{e}^{sx}}{S(x)}dx d\tau\nonumber\\
  &\quad+ \int_0^\infty \mathrm{e}^{-s \tau} P(\tau) \hat q(s,0) d\tau\nonumber \\
              &=  \hat q(s,0) \int_0^\infty \mathrm{e}^{-s \tau} P(\tau) d\tau\nonumber\\
  &\quad+  \int_0^\infty \mathrm{e}^{-s \tau} P(\tau) \int_0^\tau p(x,0) \frac{\mathrm{e}^{sx}}{S(x)}dx d\tau. \label{eq:appE_q0_eq1}
\end{align}
Equation \eqref{eq:appE_q0_eq1} can be written more compactly as
\begin{equation}
\hat q(s,0) = \frac{f(s)}{1 - P_L(s)},
\end{equation}
where we used the Laplace transform of the ISI density, $P_L(s) = \int_0^\infty \mathrm{e}^{-s \tau} P(\tau) d\tau$, and defined $f(s)$ as the second summand in Eq.~\eqref{eq:appE_q0_eq1}.
Now, $\hat q(s,0) $ has simple poles $\lambda_n$ if $P_L(\lambda_n)=1$ and $f(\lambda_n) \neq 0$.

For the particular initial condition $p(\tau,0)=\delta(\tau)$, where the population is prepared in the fully synchronized state, then the function $f(s)$ simplifies as
\begin{align}
  f(s) &= \int_0^\infty \mathrm{e}^{-s \tau} P(\tau) \int_0^\tau \delta(x) \frac{\mathrm{e}^{sx}}{S(x)}dx d\tau\nonumber\\
  &= \int_0^\infty \mathrm{e}^{-s \tau} P(\tau) d\tau = P_L(s).
\end{align}
Likewise, $\hat q(\tau,s)$ as given by Eq.\eqref{eq:appE_qhat_sol} reduces to $\hat q(\tau,s) = 1 + \hat q(0,s)$.
Thus, by Eq.~\eqref{eq:appE_qhat_def} the Laplace transform of the refractory density equation reads
\begin{equation}\label{eq:hatptau_sol1}
\hat p(\tau,s) = \mathrm{e}^{-s\tau} S(\tau) \big[ 1 + \frac{P_L(s)}{1-P_L(s)} \big] = \mathrm{e}^{-s\tau} S(\tau) \frac{1}{1-P_L(s)}.
\end{equation}
Note that $\hat p(\tau,s) $ has indeed simple poles at $s=\lambda_n$ with $P_L(\lambda_n)=1$ because $\mathrm{e}^{-\lambda_n \tau }S(\tau) \neq 0 $ for all finite $\tau>0$ and $\mathrm{Re}(\lambda_n) < \infty$.
Applying now the inverse Fourier transform to Eq.~\eqref{eq:hatptau_sol1}, yields
\begin{equation}\label{eq:appE_p_decomp_alt1}
p(\tau,t) = \sum_{n\in\mathbb{Z}} \alpha_n(\tau) \mathrm{e}^{\lambda_n t}
\end{equation}
with (possibly $\tau$-dependent) coefficients $\alpha_n$.
Equation \eqref{eq:appE_p_decomp_alt1} directly corresponds to the spectral decomposition Eq.~\eqref{eq:spec-decomp}.
What is more, Eq.~\eqref{eq:appE_p_decomp_alt1} predicts that the time-dependent components of the spectral decomposition Eq.~\eqref{eq:spec-decomp}, which are just the modes $a_n(t)$, will decay with rates $\lambda_n$, in line with Eq.~\eqref{eq:an-dyn}.
These rates $\lambda_n$ are exactly the roots of the characteristic equation \eqref{eq:condition2}.

For general initial conditions, $p(\tau,0) = \phi_0(\tau)$, a similar argument holds.
Now, we have
\begin{equation}
\hat q(\tau,s) = \hat q(0,s) + \int_0^\tau \phi_0(x) \frac{\mathrm{e}^{sx}}{S(x) }dx  =: \frac{f(s)}{1-P_L(s)} + \tilde{S}(\tau)
\end{equation}
with the function $f(s)$ as defined in Eq.~\eqref{eq:appE_q0_eq1}.
Hence, we find 
\begin{equation}\label{eq:hatptau_sol2}
\hat p(\tau,s) = \mathrm{e}^{-s\tau} S(\tau) \big[ \tilde S(\tau) + \frac{f(s)}{1-P_L(s)} \big] .
\end{equation}
Under rather general conditions, the $\lambda_n$ can be guaranteed to be simple poles of $P_L(\lambda) = 1$, resulting again in a spectral decomposition of the form Eq.~\eqref{eq:appE_p_decomp_alt1} akin to the eigenfunction expansion Eq.~\eqref{eq:spec-decomp} and characteristic equation \eqref{eq:condition2}.

\end{document}